\documentclass[trackchanges,twocolumn]{aastex701}

\usepackage{booktabs}
\usepackage{graphicx}
\usepackage{txfonts}
\usepackage{bm}
\usepackage{threeparttable}

\begin{document}

\title{The $\beta$-Dependence of Particle Spectra in Relativistic Turbulent Reconnection} 

\author[orcid=0009-0009-0818-0230,gname=Shi-Min,sname=Liang]{Shi-Min Liang}
\affiliation{School of Mathematics and Computational Sciences, Xiangtan University, Xiangtan, Hunan 411105, People’s Republic of China}
\affiliation{Department of Physics, Xiangtan University, Xiangtan, Hunan 411105, People’s Republic of China}
\email{202331510127@smail.xtu.edu.cn}  

\author{Jian-Fu Zhang} 
\affiliation{Department of Physics, Xiangtan University, Xiangtan, Hunan 411105, People’s Republic of China}
\affiliation{National Astronomical Observatories, Chinese Academy of Sciences, Beijing 100101, China}
\email[show]{jfzhang@xtu.edu.cn}

\author[orcid=0000-0002-2106-8237]{Nian-Yu Yi}
\affiliation{School of Mathematics and Computational Sciences, Xiangtan University, Xiangtan, Hunan 411105, People’s Republic of China}
\affiliation{Hunan Key Laboratory for Computation and Simulation in Science and Engineering, Xiangtan 411105, People’s Republic of China}
\email[show]{yinianyu@xtu.edu.cn}


\begin{abstract}
We perform numerical simulations of particle acceleration in relativistic, self-driven turbulent magnetic reconnection using the MHD-PIC method. We systematically investigate the dependence of the non-thermal particle spectral exponent on the plasma $\beta$. We find that particle acceleration proceeds in two stages: an initial, efficient first-order Fermi phase where momentum gains are comparable in parallel and perpendicular directions, followed by a slower drift-dominated phase. The power-law slope of the non-thermal spectrum is established during the Fermi phase, as found in previous studies. Our results demonstrate a systematic steepening of the accelerated particle energy spectrum with increasing $\beta$. We derive empirical scaling relations: the spectral exponent $\alpha \propto \beta^{0.5}$ in the relativistic regime, compared to $\alpha \propto \beta^{0.3}$ in the non-relativistic case. This marked difference is rooted in relativistic physics: the increased inertial mass density ($\rho h$) in high-$\beta$ plasmas acts as an energy sink, reducing the Alfv\'en velocity and thereby altering the dynamics of magnetic energy release and its partition efficiency. The derived scaling provides a unified physical framework for interpreting the diversity of non-thermal radiation spectra observed in astrophysical sources, including black hole corona X-ray flares, gamma-ray bursts, and active galactic nucleus jets.

\end{abstract}



\section{Introduction} \label{intro}
The study of high-energy plasma dynamics is crucial for understanding the complex phenomena associated with high-energy astrophysical sources, such as pulsars (\citealt{Uzdensky2014}), supernova remnants (\citealt{Cerutti2014b}), and active galactic nuclei (\citealt{Kadowaki2015}). These environments are characterized by extreme conditions where relativistic effects and strong electromagnetic fields dominate the plasma behavior. Observations of $\gamma$-ray sources emitting photons up to 1.4 PeV (\citealt{Cao2021}) indicate the presence of cosmic-ray factories, or PeVatrons, that accelerate particles to extreme energies. However, the precise mechanisms of particle acceleration in these environments remain poorly understood.

Magnetic reconnection is widely regarded as one of the most effective mechanisms for particle acceleration (\citealt{deGouveiadalPino2005}). This process converts magnetic field energy into plasma kinetic and thermal energy, leading to significant changes in the magnetic field topology. While the Sweet-Parker model (\citealt{Sweet1958,Parker1957}) describes steady-state reconnection, modern theories emphasize the role of turbulence and instabilities in driving fast reconnection (\citealt{Landi2015,Papini2019a,Lazarian1999}). Recent studies have shown that magnetic reconnection, particularly efficient at energizing particles through a first-order Fermi process within contracting magnetic structures. In 2D simulations, this process is often observed within magnetic islands (or plasmoids) formed by the tearing-mode instability (\citealt{Sironi2014,Drake2010}). In 3D cases, the analogous structures are flux ropes or tubes(\citealt{Guo2014,Guo2015}), and fast reconnection can be driven broadly by turbulence (\citealt{Lazarian1999}), a regime more relevant to present 3D study. The key advantage of the turbulent reconnection framework (\citealt{Lazarian1999}) is its applicability in 3D, where it does not rely on a specific instability to trigger fast reconnection but rather on the ubiquitous presence of turbulence, which efficiently drives reconnection through magnetic field line wandering (\citealt{Eyink2011}).

Over the past decade, extensive research, primarily through Particle-in-Cell (PIC) numerical simulations, has demonstrated that relativistic magnetic reconnection is a highly promising candidate for particle acceleration. Charged particles can be accelerated through a first-order Fermi process inherent to fast magnetic reconnection. This process manifests in various forms in numerical simulations: it can be described as particles scattering between converging magnetic fluxes within a reconnection layer (\citealt{deGouveiadalPino2005}), or equivalently, as acceleration during the contraction or merging of magnetic structures (often called islands in 2D or flux ropes in 3D) (\citealt{Drake2010,Kowal2011,Guo2014,Guo2015,Guo2021}). 

These findings have been extended to magnetohydrodynamic (MHD) and test-particle simulations (\citealt{Liu2009,Kowal2011,Kowal2012,Ripperda2017a,Zhang2023}), which have yielded results consistent with those of PIC simulations with respect to particle acceleration. For instance, \cite{deGouveiadalPino2005} proposed a general model for first-order Fermi acceleration within fast-reconnecting current sheets or layers. In this model, trapped particles gain energy through repeated interactions with the two converging magnetic fluxes of opposite polarity, analogous to shock acceleration. \cite{Kowal2011} later demonstrated the equivalence of this process with the acceleration observed in contracting or merging magnetic structures (islands in 2D, flux ropes in 3D) seen in PIC and MHD simulations. A key strength of this framework is its applicability in both 2D and 3D, independent of the specific mechanism that triggers the fast reconnection. In MHD simulations, the required turbulence for fast reconnection can be generated self-consistently by various instabilities, such as the current-driven kink instability (e.g., \citealt{Kadowaki2021,Medina2021,Medina2023ApJ}), the magnetic rotational instability (\citealt{Kadowaki2018ApJ}), or the Kelvin-Helmholtz instability (\citealt{Kowal2020}).

From the perspective of numerical methods, the disadvantage of PIC methods is that the influence of microdynamic scale can be observed but cannot extend to a large scale owing to the limited particle gyroradius. The MHD test particle method can study the reconnection acceleration at the macroscopic scale, which is helpful to understand the overall process on the macroscopic scale without considering the microplasma effects (e.g., \citealt{Kowal2011,Medina2021}). The MHD-PIC approach is designed to bridge the gap between kinetic-scale physics and large-scale fluid dynamics, enabling the study of particle acceleration in self-consistent, large-scale turbulent reconnection environments. 
However, a dedicated parameter study quantifying the $\beta$-dependence of the particle spectral index within a fully relativistic, self-consistent turbulent MHD-PIC framework is still needed. While studies in the non-relativistic regime have explored the dependence on Alfv\'en speed (effectively $\beta$) using MHD + test particles method (e.g., \citealt{delvalle2016}), and recent relativistic MHD-PIC simulations have begun to investigate related environments (\citealt{Medina2023ApJ,deGouveiadalPino2024arXiv}), a dedicated parameter study focusing on the $\beta$-dependence within a fully relativistic, turbulent MHD-PIC framework is still lacking. In this work, we adopt the MHD-PIC module (\citealt{Bai2015,Mignone2018}) from astrophysical simulation code PLUTO (\citealt{Mignone2007}). This hybrid MHD-PIC module is advantageous to explore the cosmic-ray (CR) dynamic effect on a scale larger than the ion’s skin depth (\citealt{Mignone2020}; see Section 2.1 for more details). 

In this work, we focus on the coevolution of particles and fluids, and investigate the role of plasma $\beta$ on particle acceleration in relativistic magnetic reconnection. This paper is organized as follows. Our simulation methods are presented in Section \ref{sec:sim_methods}. Section \ref{sec:results} describes results of the numerical simulation. In Section \ref{sec:discussion}, we provide the discussion. Finally, we present the summary in Section \ref{sec:summary}.

\section{Simulation method} \label{sec:sim_methods}
\subsection{MHD-PIC module} \label{sec:sim_MHD-PIC}
The MHD-PIC method was proposed by \cite{Bai2015}. This hybrid approach solves the MHD equations for the thermal fluid while simultaneously tracking the dynamics of CR particles using PIC technique. The module incorporates two-way coupling between the components through momentum-energy feedback and includes the CR-induced Hall term in Ohm’s law.

The MHD-PIC module embedded in the PLUTO code simulates the evolution of the fluids by solving the set of MHD equations and simultaneously integrate the motion of CR particles using conventional PIC techniques. The numerical implementation comprises three primary components: (1) a finite-volume solver for the MHD equations, (2) a particle integrator for the CR dynamics, and (3) a time-stepping scheme that maintains synchronization between the fluid and particle components. The MHD solver enforces the divergence-free condition through either constrained transport or divergence-cleaning methods. For particle integration, the module offers both time-reversible PIC algorithms and Runge-Kutta schemes. The particles represent phase-space clouds of physical particles, enabling efficient large-scale simulations while preserving kinetic effects. For details of this module, refer to \cite{Bai2015} and \cite{Mignone2018}.

\subsection{Simulation setup} \label{sec:sim_setup}
To study turbulent magnetic reconnection processes, we perform numerical simulations employing the MHD-PIC module (\citealt{Mignone2018}) mentioned above to solve the ideal MHD equations as follows:
\begin{equation}\label{eq-RMHD1}
    \frac{\partial \rho\gamma}{\partial t}+\nabla \cdot{(\gamma\rho \bm{v}_{\rm g})}=0,
\end{equation}
\begin{equation}\label{eq-RMHD2}
	\frac{\partial \bm{m}}{\partial t}+\nabla \cdot[w_t\gamma^2\bm{v}_{\rm g} \bm{v}_{\rm g} -\bm {EE}-\bm{BB}+I(p+\frac{E^2+B^2}{2})]=0,
\end{equation}
\begin{equation} \label{eq-RMHD3}
	\frac{\partial \varepsilon_t}{\partial t}+\nabla \cdot \bm m=0,
\end{equation}
\begin{equation} \label{eq-RMHD4}
	\frac{\partial \bm{B}}{\partial t}+\nabla \times \bm E=0,
\end{equation}

\begin{equation} \label{MHD-eq5}
	\nabla \cdot \bm{B}=0.
\end{equation}
These equations are the continuity, momentum, energy, induction, and solenoidal condition, respectively. Here, $\rho$ is the mass density, $\bm{m}=\gamma^2w_t \bm{v}_{\rm g}+\bm E\times \bm B$ the momentum density, $I$ unit tensor, $w_t = \rho h+B^2/\gamma^2+(\bm v_{\rm g}\cdot \bm B)^2$ the relativistic enthalpy,  $\varepsilon_t = \gamma w_{\rm t}-p+(E^2+B^2)/2$ the total energy density, $\bm{v}_{\rm g}$ the gas velocity, $p$ the gas pressure, $\bm{B}$ the magnetic field, and $\bm E=-\bm v \times \bm B$ the electric field. It should be stressed that we do not include resistive dissipation in our simulation. 

Our numerical simulations are performed in a 3D domain with physical dimensions $1.0L\times 0.5L \times 1.0L$, where the dimensionless length scale is set to $L=10^4c/\omega_{\rm p}$ with the light speed of $c=1$ and the plasma frequency of $\omega_{\rm p} = \sqrt{\rho q^2/m_{\rm p}}$ ($q$ and $m_{\rm p}$ are a charge and mass of the proton, respectively). Similar to \cite{Kowal2012} and \cite{delvalle2016}, the configuration of the initial magnetic field is considered as a Harris type by \cite{Harris1962}
\begin{equation} \label{MHD-eq6}
	\bm{B} = B_0 {\rm tanh}\frac{y}{d}\bm{e}_{\rm x},
\end{equation}
where $d$ is the initial width of the current sheet and $B_{\rm 0}=1.0$ is the initial magnetic field strength. The magnetic fields are anti-parallel to the $X$-direction, resulting in a discontinuity placed on the $X-Z$ plane. By counteracting the Lorentz force term with a thermal pressure gradient, we can obtain an initial equilibrium condition
\begin{equation} \label{MHD-eq7}
	p = \frac{(\beta +1)}{2}B_0^2-\frac{\bm B^2}{2},
\end{equation}
where $\beta$ is the plasma parameter\footnote{Note on initial equilibrium: Equation \ref{MHD-eq7}, derived from the force balance condition $\nabla(p+B^2/2) = 0$, is valid for both our relativistic and non-relativistic simulations in the initial static ($v_{\rm g}=0$) state. In the relativistic case, $p$ represents the relativistic gas pressure. We have verified that this initialization does not generate significant spurious velocities, and any minor transients are quickly overwhelmed by the development of turbulence.}, and we have considered different values of $\beta$, as shown in Table \ref{tab:parameters}. As a result, the total pressure remains invariable throughout the current sheet. In the present study, we do not include the feedback of the accelerated particles on the background plasma fields. This allows for a controlled comparison between the relativistic and non-relativistic regimes and focuses the analysis on the intrinsic dependence of the acceleration process on the plasma $\beta$. Thus, Equations (\ref{eq-RMHD1})-(\ref{eq-RMHD4}) describe the evolution of the background fluid without source terms from particles, while particles are evolved using the fields interpolated from this fluid.

The evolution of CR particles can be described by the following equations:
\begin{equation}    \label {particle-eq1} 
	\frac{{\rm d}\bm{x}_{\rm p}}{{\rm d}t} = \bm{v}_{\rm p},
\end{equation}
\begin{equation}  \label{particle-eq2}
	\frac{{\rm d}(\gamma \bm v)_{\rm p}}{{\rm d} t} =q_{\rm p}(c\bm E + {\bm v}_{\rm p}\times \bm B),
\end{equation}
where $\gamma = (1-v_{\rm p}^2/{c^2})^{-1/2}$ is the Lorentz factor, $\bm x_{\rm p}$, $\bm v_{\rm p}$ and $q_{\rm p}$ indicate the spatial position, velocity and charge to mass ratio of a charged particle (proton for our scenario), respectively. When numerically solving Equations (\ref{particle-eq1}) and (\ref{particle-eq2}) by the Boris integrator (\citealt{Boris1971}), the electric and magnetic fields $\bm E$ and $\bm B$ are computed from the magnetized fluid using a cubic spline interpolation approach.

A total of $10^6$ test particles are uniformly initialized with a Maxwellian distribution of the thermal velocity of $0.1c$ in each direction. The test particles are advanced together with the fluid using the Boris algorithm (\citealt{Mignone2018}). We use periodic boundaries in the $X$ and $Z$-directions and reflective conditions in the $Y$-direction. Although this differs from the inflow boundaries often used in magnetic reconnection setups (e.g., \citealt{Kowal2009,Kowal2020}), it is a suitable choice for our study of self-driven turbulent reconnection. In this regime, turbulent motions, rather than an imposed external inflow, self-consistently advect plasma into the reconnection region, allowing the system to evolve towards a turbulent steady state.

To numerically solve Eqs. (\ref{eq-RMHD1})-(\ref{MHD-eq5}), we choose the HLL Riemann solver with the Characteristic Tracing Contact (CT Contact) electromagnetic field averaging scheme and the second-order piecewise linear reconstruction and second-order Runge–Kutta time stepping. Except for the part on comparative studies of the resolution, we set the numerical resolution to $512\times 256\times512$ throughout this paper, leading to the grid size of $\delta L \sim 20$. To verify the rationality of our numerical resolution, we set up a high-resolution model with a resolution of $1024\times 512\times1024$ for case of $\beta =1.0$. We initialize the density to a uniform distribution $\rho=1.0$. We set the stochastic initial velocity perturbation with the maximum amplitude of $V_{\rm eps}\sim 0.1c$ to promote the generation of turbulence, that is, to expedite the development of turbulence. When the spectral energy distributions of the test particles reach a statistically steady state, we terminate the simulation at the final integration time of $t = 1.8\times 10^5 ~\omega_{\rm p}^{-1}$. 

For MHD cases, we solve non relativistic MHD equations (similar to our pervious work; \citealt{Liang2023}). The initial field distribution is identical to that in the relativistic cases (both in code unit). It should be noted that we set the Alfv\'en velocity $V_{\rm A}=1$, and speed of light $c = 10^4V_{\rm A}$, therefore, we have $1t_{\rm A}=10^4\omega_{\rm p}^{-1}$ for MHD cases\footnote{Please note that this equation only holds in non relativistic models, where the Alfv\'en velocity $V_{\rm A}=1$. However, $V_{\rm A} = B / \sqrt{\rho h + B^2}$ for RMHD case, therefore, the equivalent Alfv\'en time ($t_{\rm A} = L/V_{\rm A}$) scale discussed later is in the relativistic case, and the same code unit time ($\omega_{\rm p}^{-1}$) corresponds to different $t_{\rm A}$ for different $\beta$ models.}. The variable parameters of all models are shown in Table \ref{tab:parameters}.

\begin{table}
\caption{Variable parameters used in our simulation.} 
    \label{tab:parameters}
    \centering
    \setlength{\tabcolsep}{4pt}
    \begin{tabular}{lcccc}
    \toprule
     Name & $\beta$ & $\alpha$ & $E_{\rm cut}$\\
     \midrule
     R-$\beta$001  & 0.01 & $1.26\pm0.07$  & 19.55 \\
     R-$\beta$01   & 0.1  & $1.23\pm0.09$  & 16.06 \\
     R-$\beta$1    & 1.0  & $1.40\pm0.13$  & 7.93  \\
     R-$\beta$5    & 5.0  & $1.91\pm0.20$  & 3.28  \\
     R-$\beta$10   & 10.0 & $1.98\pm0.06$  & 1.55  \\
     R-$\beta$1-H  & 1.0  & $1.38\pm0.12$  & 5.72  \\
     \midrule
     N-$\beta$001  & 0.01 & $1.21\pm0.11$  & 39.14  \\
     N-$\beta$01   & 0.1  & $1.24\pm0.12$  & 39.11  \\
     N-$\beta$1    & 1.0  & $1.24\pm0.10$  & 23.71  \\
     N-$\beta$5    & 5.0  & $1.45\pm0.20$  & 21.59  \\
     N-$\beta$10   & 10.0 & $1.37\pm0.13$  & 13.27  \\
     \midrule
    \end{tabular}
    \begin{tablenotes}
      \item \textbf{Note.} The characters "R" and "N" in the model name stand for relativistic and non-relativistic cases, respectively. The character "H" in the model name stand for high-resolution simulation. $\beta$ represents plasma parameter; $\alpha$ represents particle energy spectral exponent; $E_{\rm cut}$ represents particle cut-off energy. The spectral index $\alpha$ and the cut-off energy $E_{\rm cut}$ are obtained from fitting the high-energy tail of the particle energy distribution at the final simulation time. Energies are in code units, $m_{\rm p} c^2$ for RMHD and $m_{\rm p}V_{\rm A}^2$ for MHD cases.
    \end{tablenotes}
\end{table}

\section{Results} \label{sec:results}
\subsection{Numerical Convergence Verification} \label{sec:results:numer}
\begin{figure*}
    \begin{flushleft}
    \hspace{0.6cm}
        \includegraphics[width=0.84\linewidth]{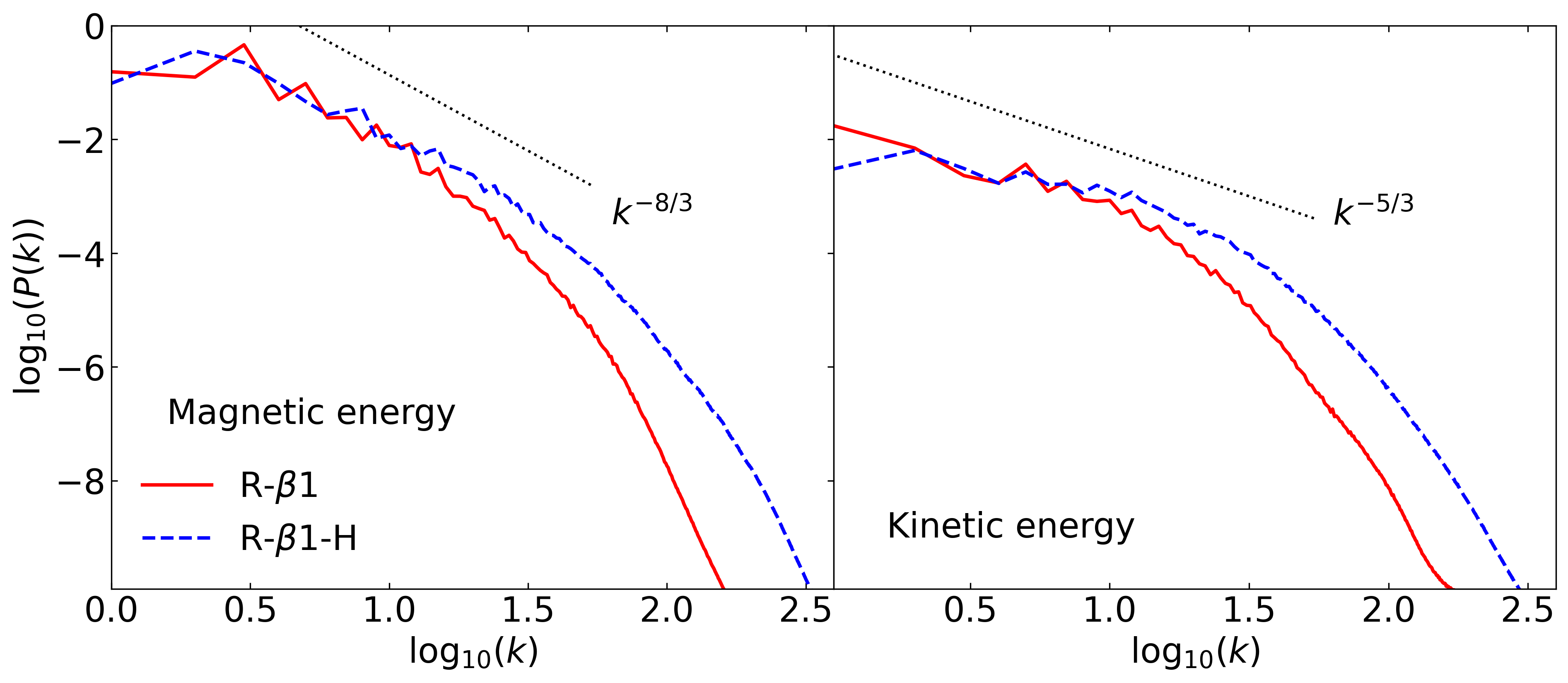}
    \end{flushleft}
    \centering
    \includegraphics[width=0.9\linewidth]{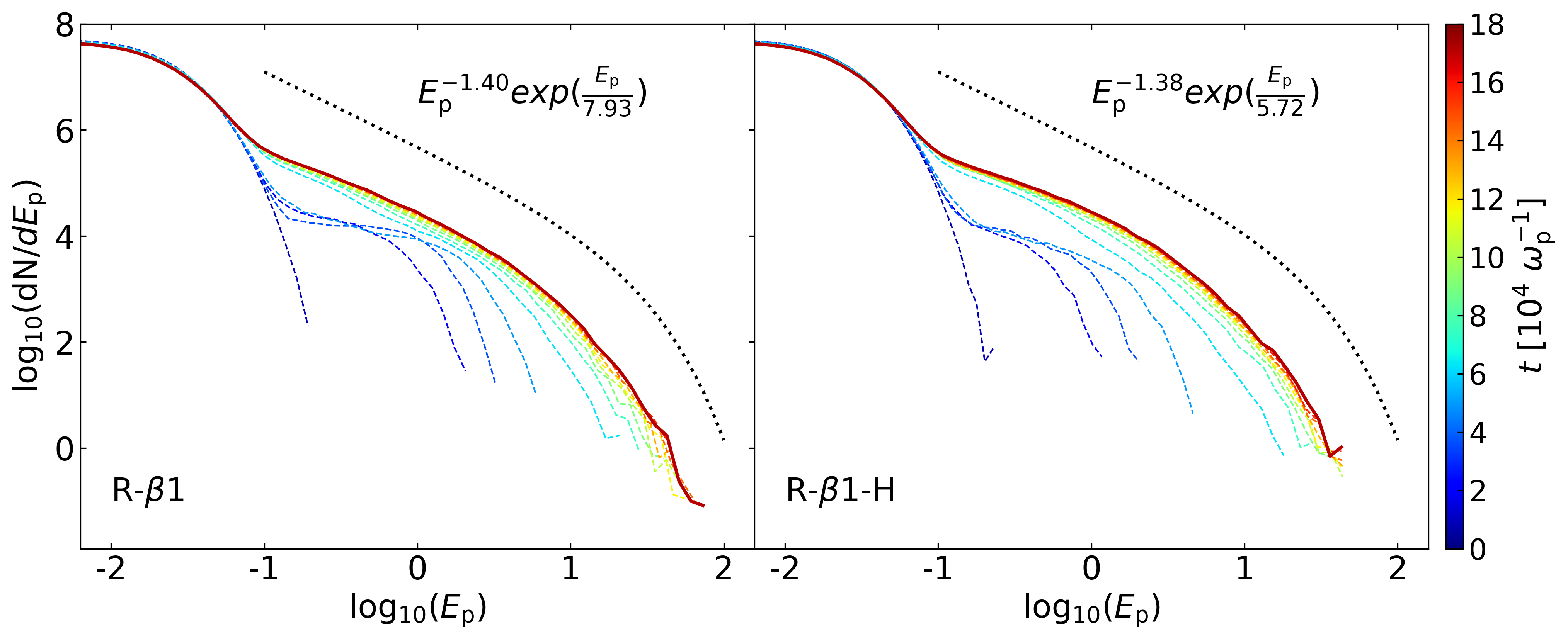}
    \caption{Power spectra (top row) of background fluids and particle energy spectra (bottom row) with different numerical resolution models at $t=1.8\times10^5\omega_{\rm p}^{-1}$. The time interval between different lines in the particle energy spectrum is $dt=1.5\times10^4\omega_{\rm p}^{-1}$. The data are based on R-$\beta1$ and R-$\beta1$-H listed in table \ref{tab:parameters}.}
    \label{fig_num_resolustion}
\end{figure*}
To ensure the reliability of our simulation results, we first performed a numerical resolution test. Figure \ref{fig_num_resolustion} shows the magnetic and velocity power spectra, as well as the final non-thermal particle energy spectra, for two different resolutions in a simulation with plasma parameter $\beta = 1$ (R-$\beta$1 and R-$\beta$1-H cases). The power spectrum is calculated in the integral form 
\begin{equation} \label{PS}
E(k_l) = L^{-1}\int \hat f(\bm k_l) \hat f^*(\bm k_l)~dk_l ~dl,    
\end{equation}
where $l$ and $k_l$ represent the integration direction ($Y$ direction) and the wave vector in the plane perpendicular to that direction, respectively, and $\hat f(\bm k_l)$ denotes the Fourier transform of $\bm v$ or $\bm B$.
This method effectively mitigates the influence of the initial anti-parallel magnetic field reversal at $y=0$ on spectral analysis (\citealt{Beresnyak2017}). The results show that the magnetic power spectrum follows an approximate $k^{-8/3}$ power law, while the velocity spectrum is close to the classical Kolmogorov $k^{-5/3}$ scaling, consistent with our previous studies of non-relativistic turbulent magnetic reconnection (\citealt{Liang2025AA}). The spectral indices are nearly identical between the two resolutions, with the higher-resolution model exhibiting only a broader inertial range. This indicates that our numerical setup is sufficient to capture the key features of the turbulent cascade within the reconnection layer. The particle energy spectra are fitted with a truncated power-law form $f(E) \propto E^{-\alpha} \exp(-E/E_{\rm cut})$, yielding spectral indices of $\alpha \approx 1.40$ (low resolution) and $\alpha \approx 1.38$ (high resolution), with comparable cutoff energies $E_{\rm cut}$. This confirms that the current numerical resolution is adequate for studying reconnection-driven particle acceleration.

Our measured magnetic and velocity power spectra in the developed turbulent state (Figure \ref{fig_num_resolustion}, top row) are consistent with classic and relativistic predictions for MHD turbulence. The magnetic spectrum scaling approximately as $k^{-8/3}$ and the velocity spectrum as $k^{-5/3}$ align with expectations for a turbulent cascade within the reconnection layer (e.g., \citealt{Beresnyak2017}). \cite{Kadowaki2021} find magnetic energy spectra with slopes between $k^{-3.3}$ and $k^{-5.1}$ and kinetic energy spectra with slopes of $k^{-3.6}$ to $k^{-3.7}$ in the fully developed turbulent regime following the saturation of the kink instability (\citealt{Medina2023ApJ} also obtained similar results). Their reported slopes encompass and are consistent with our measured values. The steeper magnetic spectra in their work are attributed to the presence of a strong guide field. This comparison indicates that, despite the different global setup, the developed turbulent state within the reconnection/dissipation region exhibits similar statistical properties, characterized by a Kolmogorov-like cascade. This reinforces the notion that the particle acceleration processes we investigate are governed by the properties of the resulting MHD turbulence, which shows robust characteristics across different driving mechanisms.

\subsection{Energy Evolution} \label{sec:results:energy}
\begin{figure*}
    \centering
    \includegraphics[width=0.96\linewidth]{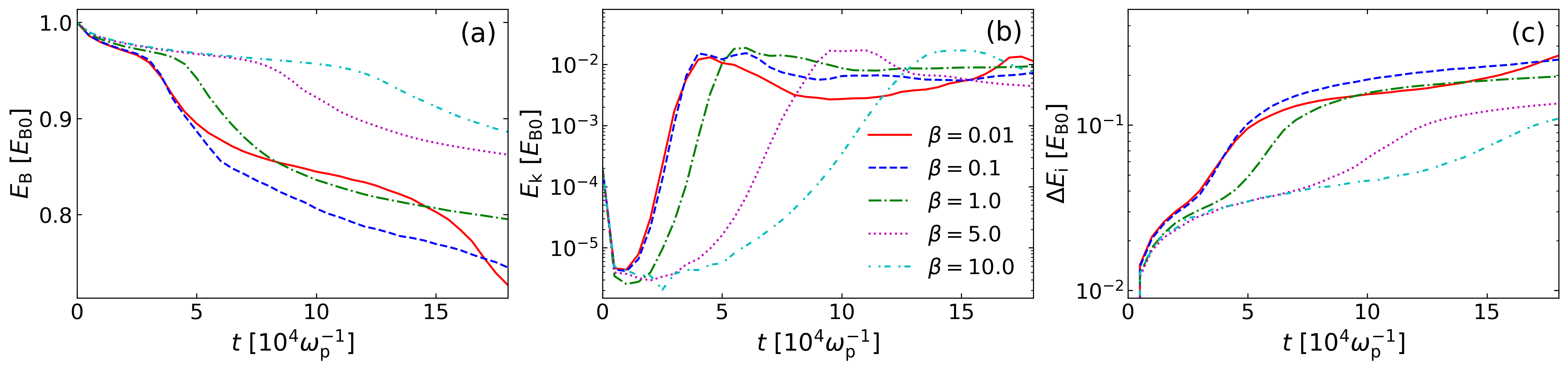}
    \includegraphics[width=0.96\linewidth]{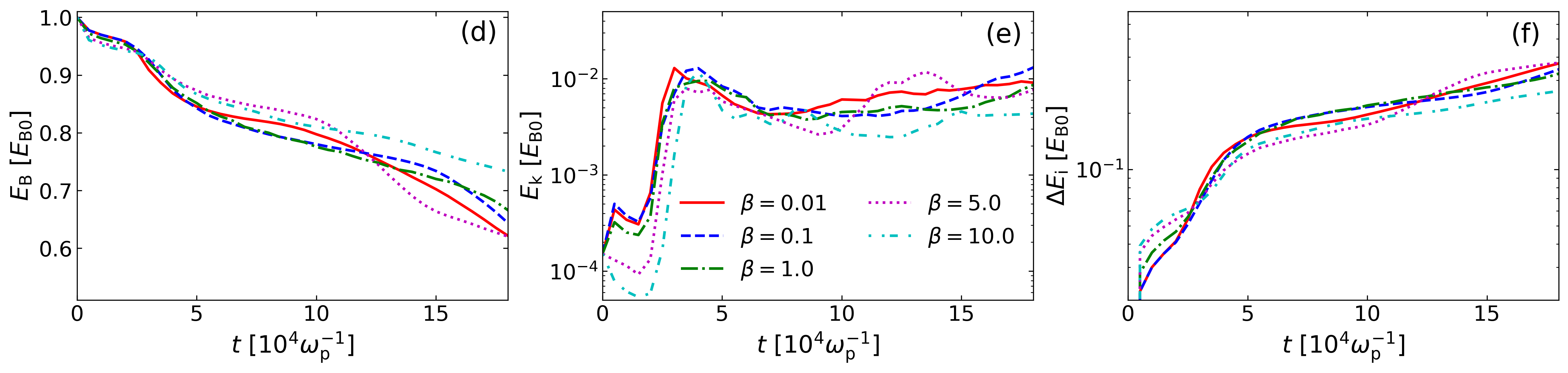}
    \caption{Magnetic energy (left column), kinetic energy (middle column) and increments of internal energy (right column) as a function of time in both RMHD (top row) and MHD (bottom row) cases. All panels are normalized to the initial magnetic energy.}
    \label{fig_energy_t}
\end{figure*}
To investigate the energy conversion mechanisms during reconnection, we analyzed the temporal evolution of global energy components. Please note that Figure \ref{fig_num_resolustion} (bottom) plots the distribution of kinetic energies of individual test particles in physical units, $m_{\rm p}c^2$, while Figure \ref{fig_energy_t} plots the volume-averaged kinetic and magnetic energy densities of the MHD fluid in code units. Since the units in the two figures are different, and their numerical values are not suitable for direct comparison. Figure \ref{fig_energy_t} displays the evolution of magnetic energy (panels (a) and (d)), kinetic energy (panels (b) and (e)), and the net increase in gas internal energy (panels (c) and (f)) for both relativistic (top row) and non-relativistic (bottom row) simulations. All panels are normalized to initial magnetic energy. As shown in Figure \ref{fig_energy_t}(a), the initial magnetic energy dissipates continuously over time. The dissipation rate and the final residual magnetic energy exhibit a strong dependence on $\beta$. For low-$\beta$ case ($\beta\ll 1$, magnetic pressure dominated), magnetic energy dissipates faster and leaves less residual energy. In contrast, for high-$\beta$ ($\beta\geq1$, thermal pressure dominated), the dissipation is more gradual. Correspondingly, the kinetic energy evolution (Figure \ref{fig_energy_t}(b)) shows that in low-$\beta$ models, the kinetic energy peaks earlier, followed by a slight decrease and eventual saturation. In high-$\beta$ models, the kinetic energy growth is more moderate, and saturation occurs later. The growth of internal energy (Figure \ref{fig_energy_t}(c)) is also negatively correlated with $\beta$, with faster growth and a larger final increase at low $\beta$. These trends are particularly pronounced for $\beta > 0.1$. 

The efficiency of magnetic energy conversion is regulated by the plasma $\beta$ through its control of the turbulent reconnection layer properties. In the low-$\beta$ regime (magnetic pressure dominated), the reduced thermal pressure allows for the development of a more dynamic and structured turbulent reconnection layer. Within this developed turbulent state, fast magnetic reconnection is enabled by the stochastic wandering and diffusion of magnetic field lines (\citealt{Lazarian1999,Kowal2009,Eyink2013,Lazarian2020,Vicentin2025,Vicentin2025arXiv}). This process efficiently brings oppositely directed magnetic fluxes into contact over a broad volume, leading to rapid magnetic energy dissipation, stronger plasma heating, and the generation of vigorous turbulent motions (kinetic energy growth). Crucially, the lower plasma inertia in this regime permits a higher effective Alfv\'en speed and thus a larger reconnection inflow velocity ($V_{\rm rec}$). Since the efficiency of the first-order Fermi acceleration process within the reconnection layer scales as $\Delta E/E \propto V_{\rm rec}$ (\citealt{deGouveiadalPino2005,Xu2023ApJ,deGouveiadalPino2024arXiv}), this results in more efficient particle energization and the harder spectra observed, as shown in Table \ref{tab:parameters} and Figure \ref{fig:alpha_beta}.

In contrast, at high $\beta$ (thermal pressure dominated), the increased thermal pressure (and in the relativistic case, the significantly increased enthalpy density $\rho h$) raises the effective inertial mass of the plasma. This reduces the Alfv\'en speed and consequently the reconnection rate $V_{\rm rec}$ at which magnetic fluxes converge. The lower $V_{\rm rec}$ directly diminishes the efficiency of the Fermi acceleration mechanism (\citealt{delvalle2016}). Furthermore, the increased pressure may also compress the inertial range of the turbulence, leading to a thinner, more quiescent current sheet where magnetic energy release and particle acceleration proceed at a slower pace. The combined effect of a reduced $V_{\rm rec}$ and altered turbulent dynamics naturally leads to the less efficient energy conversion and the systematically steeper particle energy spectra (larger $\alpha$, as shown in Table \ref{tab:parameters} and Figure \ref{fig:alpha_beta}) observed in high-$\beta$ models.

This correlation is more pronounced in relativistic simulations. This is because, in the relativistic framework, an increase in $\beta$ raises the plasma inertial mass density ($\rho h$), thereby reducing the Alfv\'en speed $V_{\rm A} = B / \sqrt{\rho h + B^2}$. Consequently, the processes of magnetic energy release and kinetic energy growth are delayed as $\beta$ increases. To verify this interpretation, we calculated the Alfv\'en time ($t_{\rm A}=L/V_{\rm A}$) and extended the simulation for the R-$\beta$10 to $4.0\times10^5 \omega_{\rm p}^{-1}$ (approximately $10.5 t_{\rm A}$). The results show that its overall evolutionary trend aligns with that of the $\beta=1$ model at $t = 1.8\times 10^5\omega_{\rm p}^{-1}$($\sim 10 t_{\rm A}$), and the peak magnetic energy dissipation rates are comparable (see Appendix \ref{appendix}), confirming the scaling of the characteristic time scales. The change in the dynamic time scale indicates that the efficiency of particle acceleration will also change, thereby affecting the final slope of the non-thermal particle energy spectrum. We will quantitatively demonstrate this impact in Section \ref{sec:results:alpha&beta}.

\subsection{Magnetic Field Geometry and Particle Acceleration Mechanisms}\label{sec:results:curv&acc}
\begin{figure}
    \centering
    \includegraphics[width=0.55\linewidth]{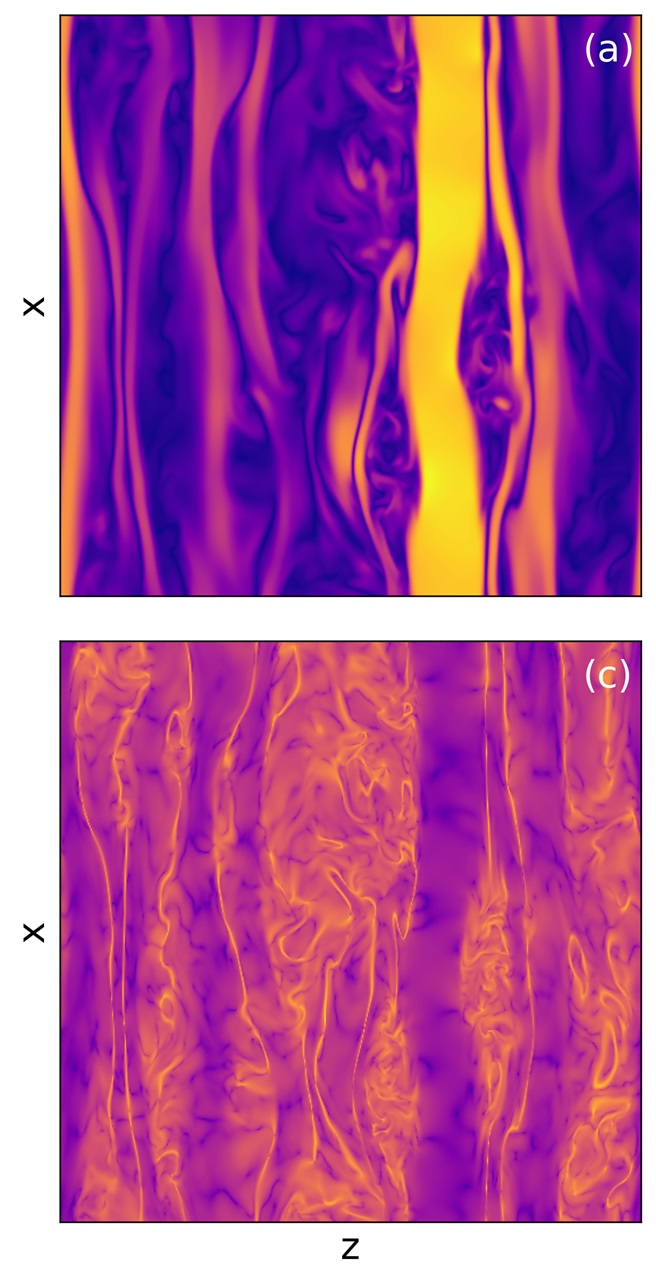}
    \includegraphics[width=0.41\linewidth]{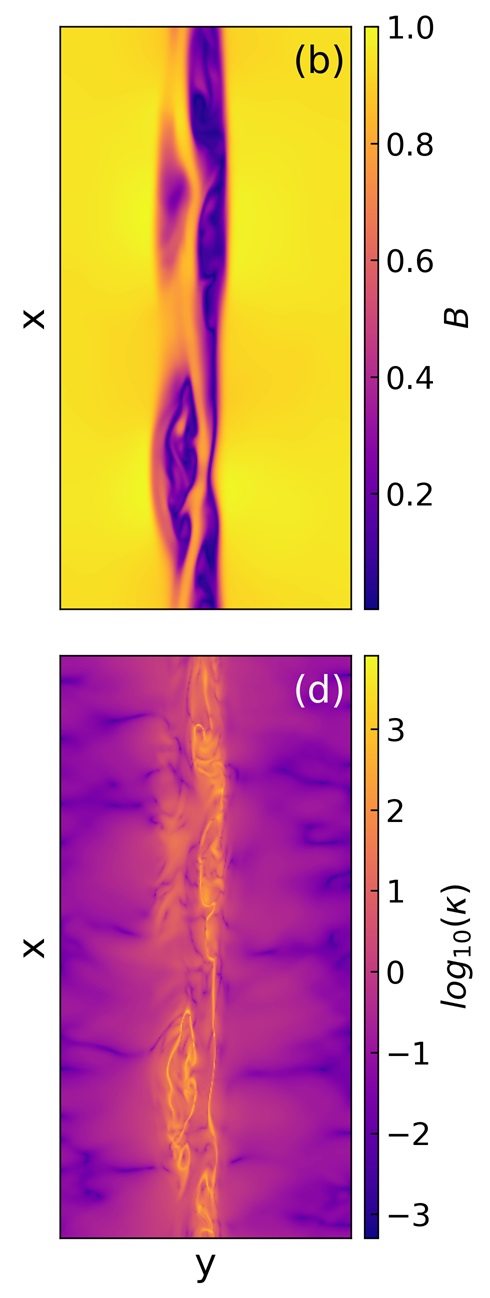}
    \caption{Magnetic field strength $B$ (top row) and field curvature $\kappa$ (bottom row) in the $xz$-plane (left column) and $xy$-plane (right column).}
    \label{fig:B_kappa}
\end{figure}
To quantify the bending of magnetic field lines within the reconnection region, we calculated the magnetic field curvature $\kappa = |\bm{b} \times (\bm{B} \cdot \nabla \bm{B})|$, where $\bm{b} = \bm{B}/B$. Figure \ref{fig:B_kappa} shows the distribution of magnetic field strength (panels (a) and (b)) and field line curvature (panels (c) and (d)) in the $xz$ ($y=0$) and $xy$-plane at time $t=1.8\times10^5 \omega_{\rm p}^{-1}$. The magnetic field structure remains largely aligned with the $x$-direction, while the curvature distribution is anti-correlated with it: regions of strong magnetic field correspond to smaller curvature, whereas regions of weak magnetic field (e.g., near reconnection X-points) exhibit larger curvature. This reflects the greater magnetic tension in strong-field regions, rendering the field lines more ``rigid". 

The spatial variation of magnetic field strength and curvature directly influences particle energization via drift motions. To quantitatively decompose the contributions of these magnetic field geometries to particle acceleration, we employ the guiding-center energy equations derived in the ideal MHD limit (\citealt{Mora2025}). In this framework, the rate of change of a particle's energy can be partitioned into two distinct terms, the curvature drift acceleration parallel to the direction of the local magnetic field 
\begin{equation} \label{eq:w_para}
    W_\parallel = \bm v_E\cdot[(\gamma v_\parallel)^2(\bm b\cdot\nabla)\bm b+\gamma^2v_\parallel(\bm v_E\cdot\nabla)\bm b],
\end{equation}
and the perpendicular gradient drift acceleration 
\begin{equation} \label{eq:w_perp}
    W_{\perp} = \frac{\mu}{m}\bm v_E\cdot\nabla(\frac{B}{\kappa}),
\end{equation}
exclusive to perpendicular-energy evolution. Where $v_{\parallel}$ represents the component of particle velocity parallel to the local magnetic field, and $v_E = \bm E \times \bm B/B^2$ is the $\bm E \times \bm B$ drift. These equations are applicable here as our test-particle simulations satisfy the guiding-center conditions (gyroradius much smaller than system scales), and the background fields are provided by an ideal MHD simulation where $E_\parallel=0$. Thus, $W_{\parallel}$ and $W_{\perp}$ (here, $W$ represents the energy gain rate normalized by the plasma frequency $\omega_{\rm p}$) together account for the total energy gain of particles, allowing us to distinguish between acceleration driven by field-line bending and by gradients in magnetic field strength.

\begin{figure*}
    \begin{flushleft}
    \hspace{0.2cm}
    \includegraphics[width=0.305\linewidth]{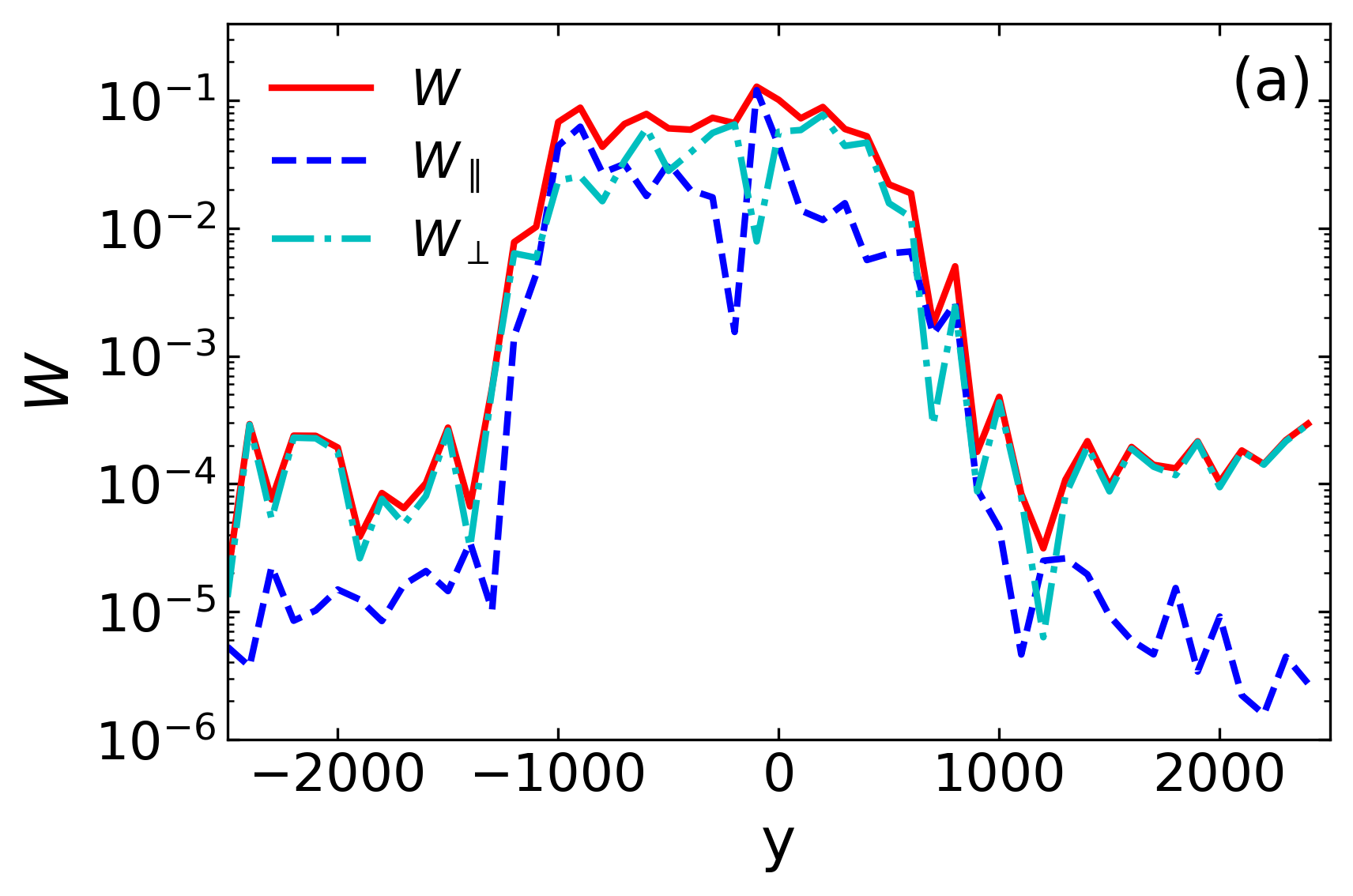}
    \includegraphics[width=0.3\linewidth]{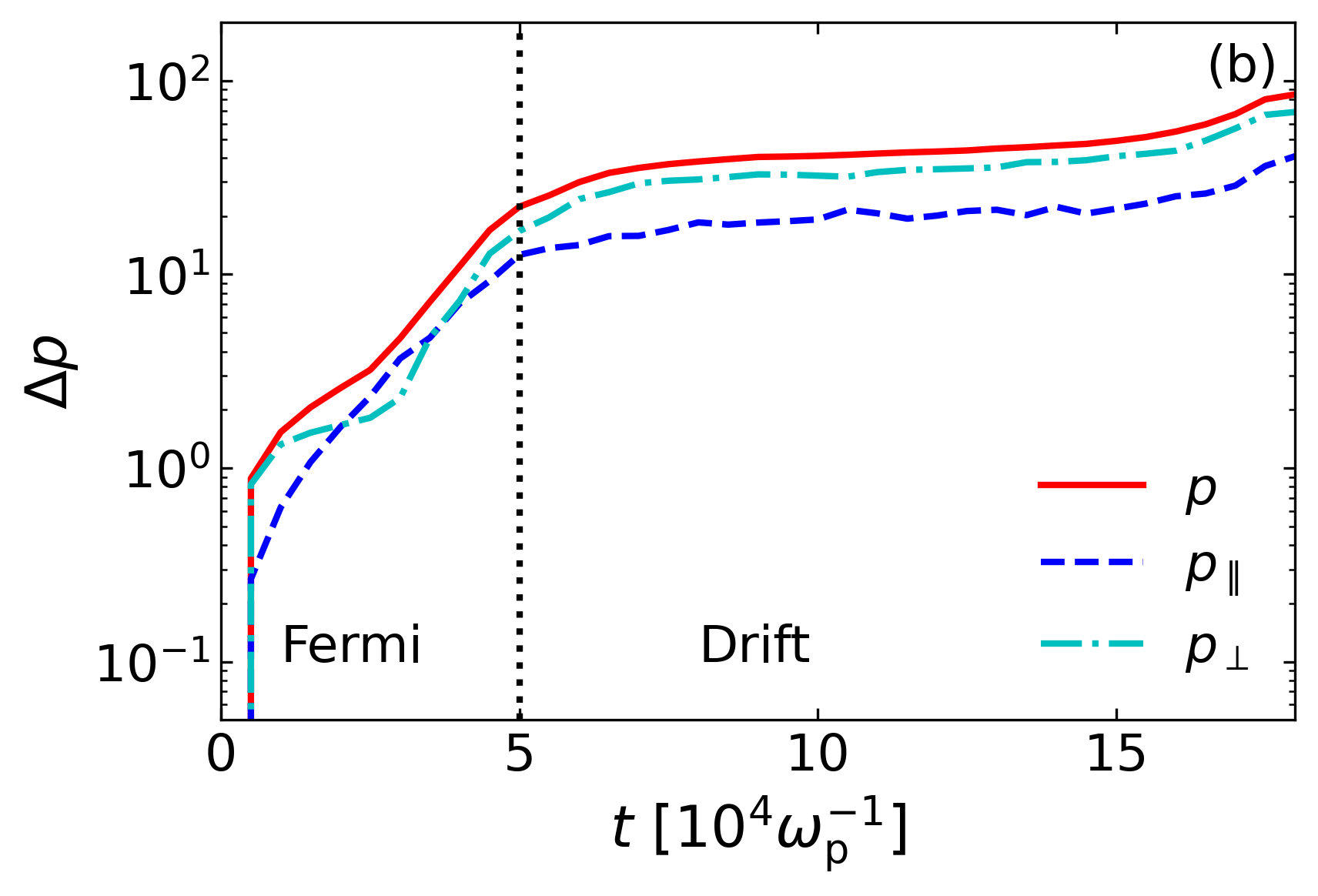}
    \hspace{0.3cm}
    \includegraphics[width=0.34\linewidth]{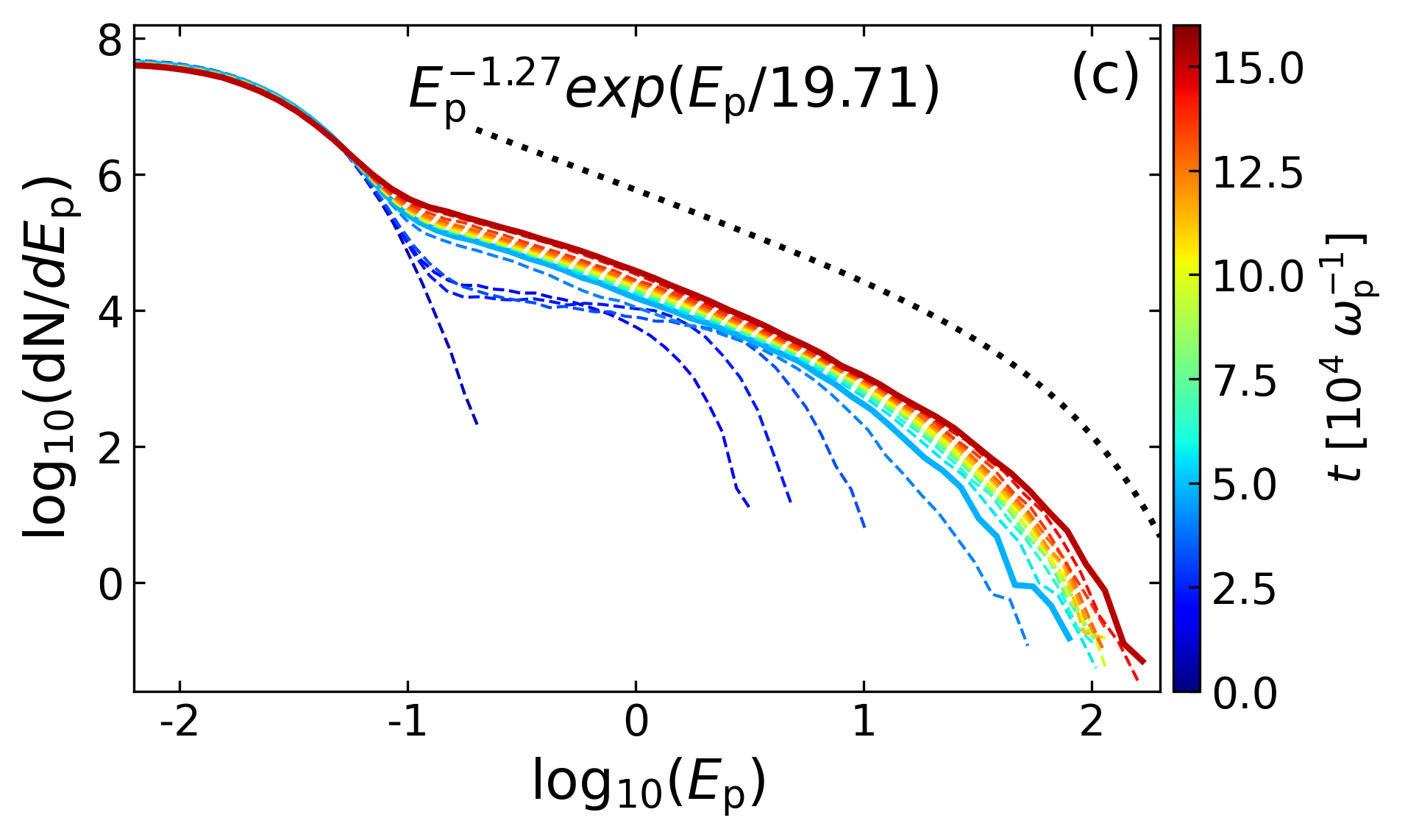}
    \end{flushleft}
    \centering
    \includegraphics[width=0.96\linewidth]{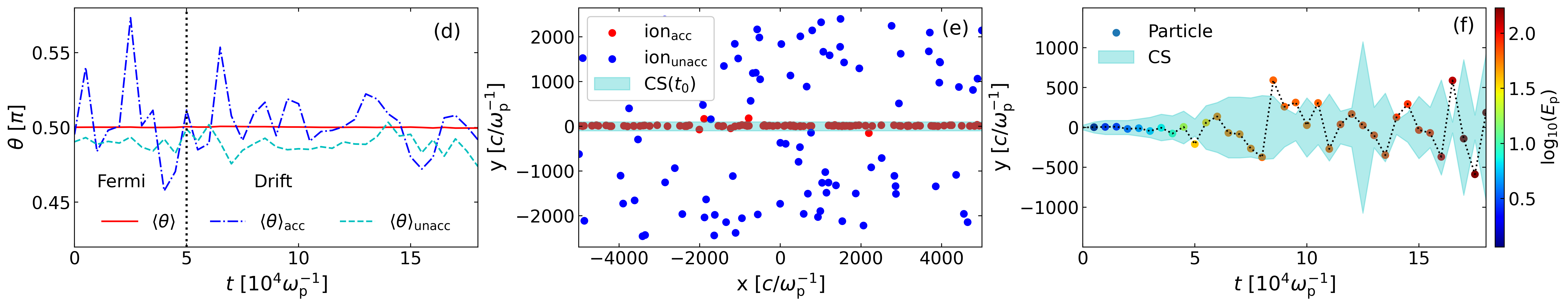}
    \caption{Panel (a): Distribution of average particle energy gain along the y-direction.
    Panel (b): The average momentum (red line) and parallel (blue dot-dashed line) and perpendicular (cyan dashed line) components of 100 accelerating particles as a function of time. The vertical dotted line represents the dividing lines of different acceleration stages, as the same as panel (d).
    Panel (c): The energy spectra of all particles at different evolutionary times (color bar). The cyan and red solid lines represent the energy spectra at $t=5\times10^4\omega_{\rm p}^{-1}$ and $t=1.8\times10^5\omega_{\rm p}^{-1}$.
    Panel (d): The pitch-angle of all particles (red line), 100 accelerated particles (blue dot-dashed line) and 100 non-accelerated particles (cyan dashed line) as a function of time, the vertical dashed line represents the transition between Fermi and Drift stages. 
    Panel (e): The initial position distribution of 100 accelerated particles (red point) and 100 non-accelerating particles (blue point) on the x-y plane. The light cyan shaded area represents the initial width of the current sheet.
    Panel (f): The y-direction coordinate of a selected accelerated particle as a function of time. The color of the scatter represents particle energy, the light cyan shaded area represents the width of the current sheet at the position of the particle at the corresponding time. All panels are based on R-$\beta$001.}
    \label{fig:particle_dynamics}
\end{figure*}
Based on Equations \ref{eq:w_para} and \ref{eq:w_perp}, we further analyzed particle behavior using the R-$\beta$001 case. We present distribution of average particle energy gain rate $W$ along the y-direction at $t=1.8\times10^5\omega_{\rm p}^{-1}$, as shown in Figure \ref{fig:particle_dynamics}(a). The particle energy gain rate is highest near $y=0$, and the energy gain rates in the perpendicular and parallel directions are comparable, corresponding to isotropic Fermi acceleration stage in the current sheet (\citealt{Kowal2012,delvalle2016,Medina2021}). When $|y| \geq 1000 $, perpendicular acceleration begins to dominate, while parallel acceleration is much smaller than perpendicular acceleration, which may correspond to drift acceleration dominated by the perpendicular acceleration outside the current sheet.

We analyze the time evolution of the momentum gain, $\Delta p = p - p_0$, for a sample of 100 accelerated particles. As shown in Figure \ref{fig:particle_dynamics}(b), their average momentum increases rapidly for $t \lesssim 5.0 \times 10^4\omega_{\rm p}^{-1}$, after which it saturates. During this initial phase, the gains in perpendicular ($\Delta p_\perp$) and parallel ($\Delta p_\parallel$) momentum are comparable, consistent with an isotropic, first-order Fermi acceleration process inside the turbulent reconnection layer (\citealt{Kowal2012,delvalle2016}). For $t \gtrsim 5 \times 10^4\omega_{\rm p}^{-1}$, $\Delta p_\perp$ gradually exceeds $\Delta p_\parallel$, indicating a transition to a drift-dominated acceleration regime. This occurs once a particle’s Larmor radius exceeds the thickness of the current sheet, allowing it to escape the main reconnection region. Subsequent energy gain is then primarily perpendicular to the magnetic field and proceeds at a much lower rate than the earlier Fermi stage.

Our results align with the two-stage acceleration picture seen in previous studies of turbulent reconnection (\citealt{Kowal2012,delvalle2016,Medina2021,Medina2023ApJ,deGouveiadalPino2024arXiv}). The initial rapid growth in average momentum (panel (b) for $t \lesssim 5 \times 10^4 \omega_{\rm p}^{-1}$) corresponds to the first-order Fermi acceleration within the turbulent reconnection layer, where particles gain energy exponentially. During this stage, the gains in parallel and perpendicular momentum are comparable, leading to a near-isotropic distribution (\citealt{Kowal2012,Medina2021,Medina2023ApJ}). Once a particle’s Larmor radius exceeds the thickness of the reconnection layer, it escapes the efficient Fermi acceleration region and enters a drift-dominated regime (\citealt{delvalle2016,Medina2021,Medina2023ApJ,deGouveiadalPino2024arXiv}). In this later stage ($t \gtrsim 5\times10^4 \omega_{\rm p}^{-1}$), further acceleration is very slow and primarily contributes to perpendicular momentum (panel (b)), but adds little to the total energy gain. Therefore, while the time-averaged momentum gain appears dominated by the perpendicular component, the spectral shape is primarily determined by the efficient, more isotropic Fermi acceleration phase (as indicated by the cyan solid line in panel (c)).

Figure \ref{fig:particle_dynamics}(d) shows the evolution of the particle pitch-angle relative to the local magnetic field. The average pitch-angle for the entire particle population (red line) remains around $\pi/2$, reflecting the initial isotropic injection. A vertical dotted line at $t \simeq 5\times10^4 \omega_{\rm p}^{-1}$ marks the transition between the two acceleration regimes identified in Figure \ref{fig:particle_dynamics}(b) (corresponding to the Fermi-drift transition shown in panel (b)). The pitch-angles of accelerated particles (blue dot-dashed line) exhibit significant stochastic fluctuations during the Fermi acceleration regime ($t \leq 5.0 \times 10^4 \omega_p^{-1}$), a signature of the violent electromagnetic fluctuations within the turbulent reconnection layer where they are being energized. These fluctuations persist during the subsequent drift regime, but their amplitude slightly decreases. In contrast, the pitch-angles of non-accelerated particles (cyan dashed line) remain steady throughout both regimes. This comparison confirms that the intense pitch-angle scattering is directly associated with the active acceleration process within the turbulent layer and is not a feature of the background plasma.

Figure \ref{fig:particle_dynamics}(e) shows the initial positions in the $xy$-plane for the 100 highest-energy particles (red points) and the 100 lowest-energy particles (blue points) at the end of the simulation. The top 100 particles with the highest energy mainly come from within the initial current sheet, rather than accelerating particles uniformly distributed outside the current sheet. The physical origin of the Fermi-Drift transition is verified in Figure \ref{fig:particle_dynamics}(f). We track a representative accelerated particle (color-coded by energy) and estimate the local current sheet thickness (cyan shading) using the criterion from \cite{Kadowaki2021}. At the transition time $t \approx 5.0 \times 10^4 \omega_{\rm p}^{-1}$, the particle's Larmor radius ($r_{\rm L} \approx 372.3 c/\omega_{\rm p}$) becomes of the order of the maximum current sheet thickness. This confirms that the saturation of the exponential Fermi phase occurs when $r_L$ becomes comparable to the maximum scale of the turbulent reconnection layer that characterizes the scale of the acceleration region (see also \citealt{deGouveiadalPino2024arXiv}). Beyond this point, particles can no longer be efficiently confined within the reconnecting layer; they escape and undergo slower, energy-dependent drift acceleration in the large-scale, non-reconnecting magnetic fields, which is predominantly perpendicular, as seen in Figure \ref{fig:particle_dynamics}(b).

\subsection{Dependence of Particle Spectral Exponent on Plasma $\beta$} \label{sec:results:alpha&beta}
\begin{figure}
    \centering
    \begin{flushleft}
    \hspace{0.8cm}
    \includegraphics[width=0.85\linewidth]{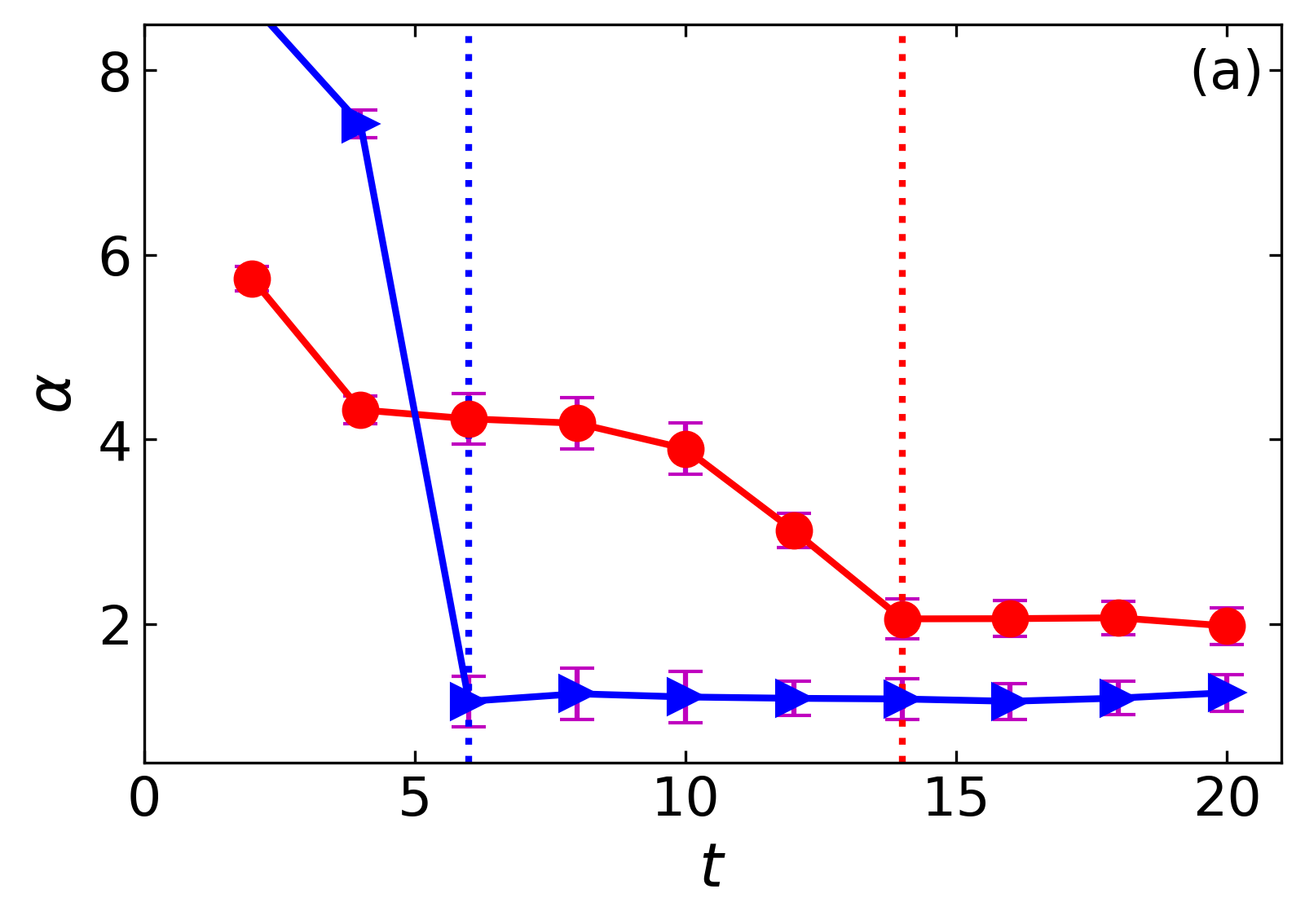}
    \end{flushleft}
    \includegraphics[width=0.9\linewidth]{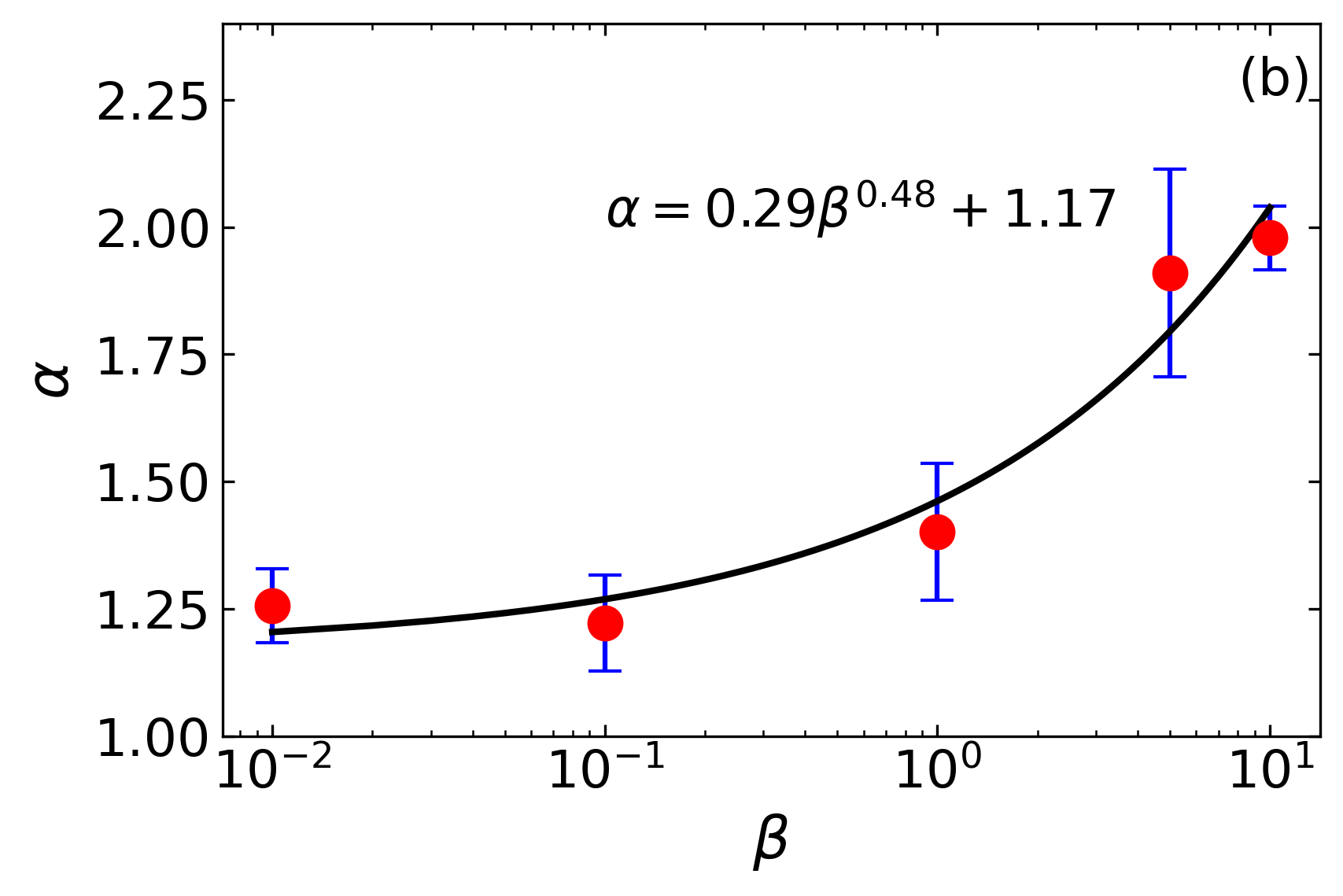}
    \includegraphics[width=0.9\linewidth]{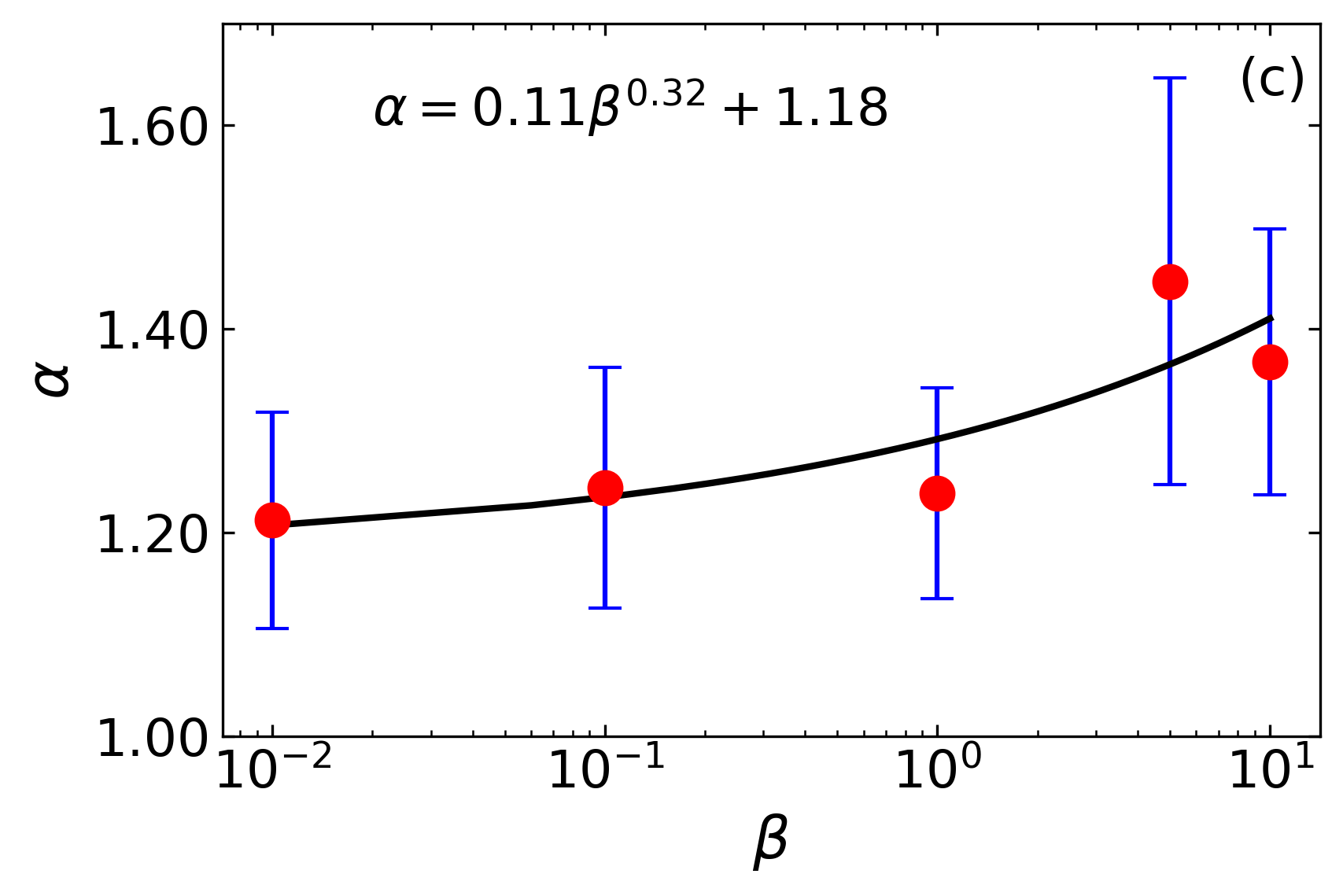}
    \caption{Panel (a): Particle spectral exponent $\alpha$ as a function of time for R-$\beta$10 case. The vertical dotted line represents the dividing lines of different acceleration stages, as the same as panel (b) of Figure \ref{fig:particle_dynamics}.
    Panels (b) and (c): Particle spectral exponent $\alpha$ as a function of plasma parameter $\beta$ in both relativistic (panel (b)) and non-relativistic (panel (c)) cases. 
    }
    \label{fig:alpha_beta}
\end{figure}

The central objective of this study is to determine the dependence of the non-thermal particle spectral exponent $\alpha$ (defined by $f(E) \propto E^{-\alpha} \exp(-E/E_{\rm cut})$) on the plasma beta $\beta$ in relativistic turbulent magnetic reconnection. Table \ref{tab:parameters} lists the fitted spectral exponent $\alpha$ and cutoff energy $E_{\rm cut}$ at the final simulation time for all models. 
The time evolution of the spectral index $\alpha$ for representative low-$\beta$($\beta = 0.01$) and high-$\beta$ ($\beta=10.0$) cases is shown in the new Figure \ref{fig:alpha_beta}(a). The index rapidly converges to a steady value during the initial Fermi acceleration phase ($t \lesssim 5 \times 10^4 \omega_p^{-1}$ for R-$\beta$001 and $t \lesssim 1.4 \times 10^5 \omega_p^{-1}$ for R-$\beta$10) and remains stable thereafter. This confirms that the power-law slope is established during the period of efficient first-order Fermi acceleration within the turbulent reconnection layer, before particles saturate and transition to the drift regime (also see \citealt{delvalle2016,Medina2021,Medina2023ApJ}). The highest-energy part of the steady-state spectrum may be influenced by the slow continued energization and recirculation of saturated particles through the periodic boundaries.

The systematic trend of steeper spectra (larger $\alpha$) with increasing $\beta$, shown in Figure \ref{fig:alpha_beta}(b), arises from the $\beta$-dependence of the turbulent reconnection process. In the high-$\beta$ regime, the increased thermal pressure (and in the relativistic case, the significantly larger enthalpy density $\rho h$) raises the effective inertia of the plasma. This reduces the Alfv\'en speed $V_{\rm A}$ and consequently the reconnection inflow velocity $V_{\rm rec}$, which scales with $V_{\rm A}$. Since the efficiency of the first-order Fermi acceleration mechanism scales as $\Delta E/E \propto V_{\rm rec}$ (\citealt{deGouveiadalPino2005,Xu2023ApJ,deGouveiadalPino2024arXiv}), a lower $V_{\rm rec}$ directly leads to less efficient particle energization and a steeper steady-state spectrum. This physical explanation and the observed $\alpha-\beta$ trend are consistent with earlier findings in non-relativistic turbulent reconnection studies (e.g., \citealt{delvalle2016}).

The empirical relations obtained from fitting are: $\alpha = 0.29\beta^{0.48}+1.17$ (RMHD case) and $\alpha = 0.11\beta^{0.32}+1.18$ (MHD case). For $\beta \ll 1$ (magnetic pressure dominated), in both relativistic and non-relativistic cases, the spectral exponent $\alpha$ remains stable in a relatively flat range ($\sim 1.2-1.25$). For the relativistic cases, $\alpha$ increases rapidly with $\beta$, reaching $\alpha \approx 2.0$ at $\beta=10$. In contrast, the increase in the non-relativistic model is much weaker (Figure \ref{fig:alpha_beta}(c)), with $\alpha \approx 1.4$ at $\beta=10$. This divergence stems from relativistic effects: for RMHD case, the Alfv\'en speed is given by $V_{\rm A}=B/\sqrt{\rho h+B^2}$, where $h$ is the specific enthalpy. In high-$\beta$ plasmas, h can become very large, especially at relativistic temperatures. This significantly increases the denominator and reduces $V_{\rm A}$ more drastically than in the non-relativistic case. This enhanced sensitivity of $V_{\rm A}$ to $\beta$ in turn increases the sensitivity of the reconnection rate $V_{\rm rec}$, and therefore affects particle acceleration.
  
\section{Discussion} \label{sec:discussion}
\subsection{The impact of numerical settings}
Using the MHD-PIC method, we have investigated particle acceleration in self-driven, relativistic turbulent magnetic reconnection. Our simulations reveal a clear and quantifiable dependence of the non-thermal particle spectral exponent $\alpha$ on the plasma $\beta$. The synthesized results indicate that the acceleration process is modulated by the plasma magnetization and is fundamentally governed by relativistic effects, operating within a two-stage framework consistent with earlier studies of turbulent reconnection (\citealt{Kowal2012,delvalle2016,Kadowaki2021,Medina2021,Medina2023ApJ,deGouveiadalPino2024arXiv}).

We employ periodic boundary conditions in the $X$- and $Z$-directions, particles that cross these boundaries re-enter the computational domain. This means particles can circulate through the reconnection layer multiple times, potentially gaining energy beyond what might be achievable in a single traversal. This could influence the long-term saturation of the acceleration process. However, the primary focus of this study is the shape of the non-thermal particle energy spectrum (i.e., the power-law exponent $\alpha$) during the statistically steady state of the turbulent reconnection layer. As demonstrated in Figure \ref{fig:mom_spectra} (Appendix A) for the high-$\beta$ case (R-$\beta$10), extending the simulation time by more than a factor of two leads to no significant change in the fitted spectral index $\alpha$ or the cutoff energy $E_{\rm cut}$. This indicates that the system reaches a quasi-steady energy distribution within a few Alfv\'n times. The scaling relation between $\alpha$ and $\beta$ presented in Section \ref{sec:results:alpha&beta} is derived from this statistically steady stage. Therefore, while periodic boundaries permit particle recirculation, they do not alter our conclusion regarding how $\beta$ regulates the steady-state acceleration efficiency and the resulting non-thermal spectral slope.

\subsection{The Two-Stage Acceleration Scenario}
The temporal evolution of particle momentum (Figure \ref{fig:particle_dynamics}b) and global energies (Figure \ref{fig_energy_t}) aligns with the established picture of acceleration in turbulent reconnection layers. Particles experience an initial phase of rapid, exponential energy growth for $t \lesssim 5 t_{\rm A}$. During this first-order Fermi-dominated stage, particles are efficiently energized within the turbulent current sheet by scattering between converging magnetic fluxes, with comparable gains in parallel and perpendicular momentum (\citealt{deGouveiadalPino2005,Kowal2012}). Subsequently, once a particle’s Larmor radius exceeds the maximum thickness of the turbulent-reconnection layer, it escapes the primary acceleration zone and enters a drift-dominated stage ($t \gtrsim 5 t_{\rm A}$). Here, further energization is slow and primarily contributes to perpendicular momentum, adding little to the total energy gain (\citealt{delvalle2016,deGouveiadalPino2024arXiv}). This explains why the time-averaged momentum gain appears dominated by the perpendicular component (Figure \ref{fig:particle_dynamics}b), while the critical spectral shape is established during the earlier, more isotropic Fermi phase.

The global energy evolution (Figure \ref{fig_energy_t}) shows that higher $\beta$ (i.e., greater thermal pressure) slows the dissipation of magnetic energy and the growth of kinetic energy. This is a direct consequence of the reduced effective Alfv\'en speed. In the relativistic regime, an increase in $\beta$ significantly raises the specific enthalpy $h$, thereby increasing the inertial mass density $\rho h$ and reducing $V_{\rm A}$. A lower $V_{\rm A}$ implies a lower reconnection rate $V_{\rm rec}$, which governs the efficiency of the first-order Fermi process ($\dot{E}/E \propto V_{rec}/c$). Therefore, the dynamics of energy conversion are intrinsically slower in high-$\beta$ relativistic plasmas.

\subsection{The $\beta$-Dependence of the Spectral Index}
The central result of this work is the empirical scaling of the particle spectral exponent with plasma $\beta$: $\alpha \propto \beta^{0.5}$ for RMHD and $\alpha \propto \beta^{0.3}$ for MHD (Figure \ref{fig:alpha_beta}). This quantifies a trend anticipated in earlier studies. \cite{delvalle2016} found in non-relativistic MHD simulations that the acceleration efficiency (and thus spectral hardness) decreased with decreasing Alfv\'en speed ($t_{\rm acc}\propto (V_{\rm A}/c)^{-\kappa}$ with $\kappa\sim2.1-2.4$), which is effectively equivalent to increasing $\beta$ for a fixed magnetic field strength. Our non-relativistic results are consistent with this finding.

The scaling of the spectral index with $\beta$ reported here is qualitatively similar to that found in non-relativistic studies (\citealt{delvalle2016}). The key advance here is the demonstration of a markedly stronger $\beta$-dependence in the relativistic case. This heightened sensitivity stems from relativistic thermodynamics. A quantitative difference may arise from relativistic effects, which modify the expression for the Alfv\'en speed to $V_A = B / \sqrt{\rho h + B^2}$. In high-$\beta$, high-temperature plasmas, the large specific enthalpy $h$ can make the plasma effectively more inert, potentially enhancing the reduction of $V_{\rm A}$ and $V_{\rm rec}$ with increasing $\beta$ compared to the non-relativistic case. This explanation is consistent with the study of \cite{Takamoto2015ApJ}, who showed that compressible turbulence effects modify the reconnection rate in relativistic plasmas.

\subsection{Comparison with Related Studies and Astrophysical Implications}
The particle spectral exponent in our low-$\beta$ models is $\alpha \sim 1.2-1.3$. This is consistent with theoretical expectations for strong magnetization (\citealt{Guo2014,Guo2015}) and with results from 3D PIC simulations in highly magnetized regimes (\citealt{Werner2016ApJL,Guo2021}). 
This result is consistent with the hard spectra ($\alpha \sim 1.2$) found in 3D relativistic MHD+test particles and MHD-PIC simulations of turbulent reconnection in jets, in the early regime of Fermi acceleration (\citealt{Medina2021,Medina2023ApJ}). However, it is lower than the spectral index reported for turbulent reconnection in MHD + test-particle simulations (\citealt{Zhang2023}). In addition, 2D PIC and MHD-PIC simulations of reconnection often yield softer spectra ($\alpha \geq 2$, \citealt{Ball2018ApJ,Werner2018MNRAS,Liang2025PoP}). This discrepancy between 2D and 3D outcomes highlights the significant role of spatial dimensionality, likely due to differences in particle confinement and the available turbulent wave modes in a full 3D geometry.

The scaling relations derived from our model provide a unified, parameter-based framework for interpreting the diversity of non-thermal radiation spectra from high-energy astrophysical sources. In strongly magnetized environments ($\beta \ll 1$) such as black hole coronae, our model predicts hard spectra ($\alpha \sim 1.2–1.3$), consistent with the hard power-law tails observed in X-ray flares (\citealt{Remillard2006ARAA,Fabian2015MNRAS}). For gamma-ray bursts, where the plasma $\beta$ in internal shocks or the jet interior may be of order unity or higher, the model yields a range of spectral indices from harder ($\alpha \sim 1.5$) to shock-like ($\alpha \sim 2.0$) values, comfortably encompassing typical observed indices (\citealt{Kumar2015,Nava2011AA}). 

Our finding that lower-$\beta$ environments produce harder particle spectra provides a natural explanation for the observed hard gamma-ray spectra in highly magnetized astrophysical sources, as it implies more efficient particle acceleration. The Fermi acceleration within a 3D turbulent reconnection layer provides a viable mechanism for in situ particle energization in such jets. This picture is consistent with, and extends, previous applications of turbulent reconnection models to AGN jet emission (e.g., \citealt{Kadowaki2015,Singh2015ApJ}). Notably, recent studies have specifically applied this framework to model the multi-wavelength spectral energy distributions and the production of very-high-energy gamma-rays and neutrinos in blazars, demonstrating its capability to explain key observational features (\citealt{Rodriguez-Ramirez2019ApJ,deGouveiaDalPino2025MNRAS}). Our results (along with earlier results of \citealt{Medina2021,Medina2023ApJ} and \citealt{deGouveiadalPino2024arXiv}) provide the underlying plasma-physical basis for the acceleration efficiency assumed in such phenomenological models. Therefore, when interpreting emission from such sources, comparisons should be made with reconnection-based acceleration studies (e.g., \citealt{Sironi2015MNRAS,Christie2019MNRAS,Medina2021,Medina2023ApJ}) in addition to shock models. 

\section{Summary} \label{sec:summary}
We have conducted simulations of relativistic magnetic reconnection and associated particle acceleration using the MHD-PIC method, which self-consistently evolves both the fluid and kinetic components. Within this framework, we systematically investigated the dependence of the accelerated particle spectral exponent on the plasma parameter $\beta$. Our principal findings are summarized as follows:
\begin{enumerate}
    \item The steady-state energy spectrum of accelerated particles is well-fitted by a power law whose index $\alpha$ systematically increases (spectrum steepens) with plasma $\beta$. This dependence originates from the reduction of the Alfv\'en speed $V_{\rm A}$ with increasing $\beta$ (e.g., from $\sim0.70c$ at $\beta = 0.01$ to $\sim 0.26c$ at $\beta = 10$ in our models). A lower $V_{\rm A}$ leads to a lower reconnection inflow speed $V_{\rm rec}$, which diminishes the efficiency of the first-order Fermi acceleration process. This quantitative link between spectral hardening and higher $V_{\rm A}$ (or lower $\beta$) is consistent with prior studies of turbulent reconnection (e.g., \citealt{delvalle2016}).
    \item Particle acceleration proceeds in two distinct regimes. An initial phase of rapid, exponential energy growth is consistent with first-order Fermi acceleration within the turbulent reconnection layer (see also \citealt{Kowal2012,delvalle2016,Medina2021,Medina2023ApJ,deGouveiadalPino2024arXiv}). In this stage, particles gain energy exponentially with comparable increases in parallel and perpendicular momentum ($\Delta p_{\parallel} \simeq \Delta p_{\perp}$). This phase sets the power-law slope of the energy spectrum. Subsequently, when a particle's Larmor radius exceeds the thickness of the turbulent layer, particles enter a drift-dominated stage where further energization is slow and primarily perpendicular ($\Delta p_{\perp} > \Delta p_{\parallel}$), contributing little to the overall spectral shape.
    \item The accelerated particle energy spectrum systematically steepens ($\alpha$ increases) with increasing $\beta$ in both regimes. We derive empirical scaling relations: $\alpha \propto \beta^{0.5}$ for RMHD cases and $\alpha \propto \beta^{0.3}$ for MHD cases. This heightened sensitivity in relativistic plasmas is a direct consequence of the relativistic enthalpy effect. 
    \item The three-dimensional turbulent environment is crucial. It facilitates fast reconnection through magnetic field line wandering and creates a volume-filling network of acceleration regions, contrasting with the island-confined acceleration often seen in 2D simulations (see also \citealt{deGouveiadalPino2024arXiv}).
    
\end{enumerate}

\begin{acknowledgments}
This work is supported in part by the High-Performance Computing Platform of Xiangtan University. J.F.Z. is grateful for the support from the National Natural Science Foundation of China (No. 12473046), and the Hunan Natural Science Foundation for Distinguished Young Scholars (No. 2023JJ10039). N.Y.Y. is grateful for the support from the National Natural Science Foundation of China (No. 12431014) and the Project
of Scientific Research Fund of the Hunan Provincial Science and Technology Department (2024ZL5017). S.M.L. is grateful for the support from the Xiangtan University Innovation Foundation for Post-graduate Students (No. XDCX2024Y164).
\end{acknowledgments}

\software{PLUTO \citep{Mignone2007,Mignone2018}}

\appendix

\section{Long-term Evolution of High-$\beta$ Model}
\label{appendix}
\begin{figure}
    \centering
    \includegraphics[width=0.6\linewidth]{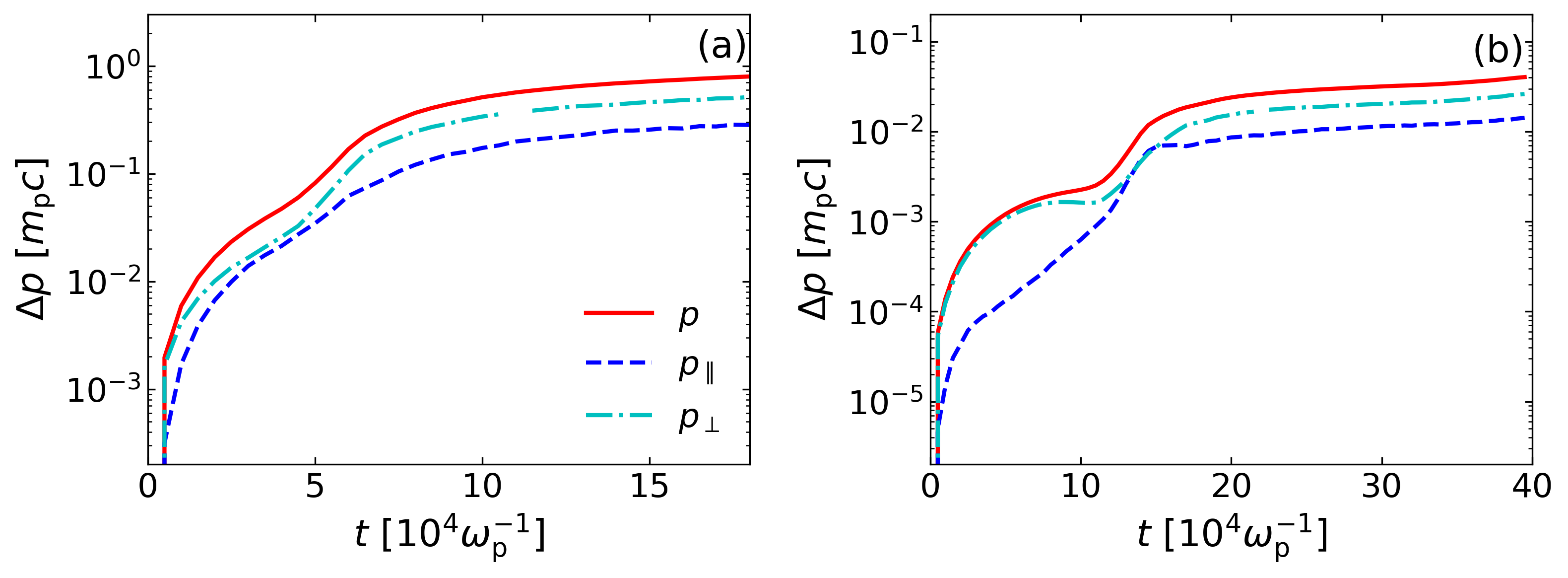}
    \includegraphics[width=0.32\linewidth]{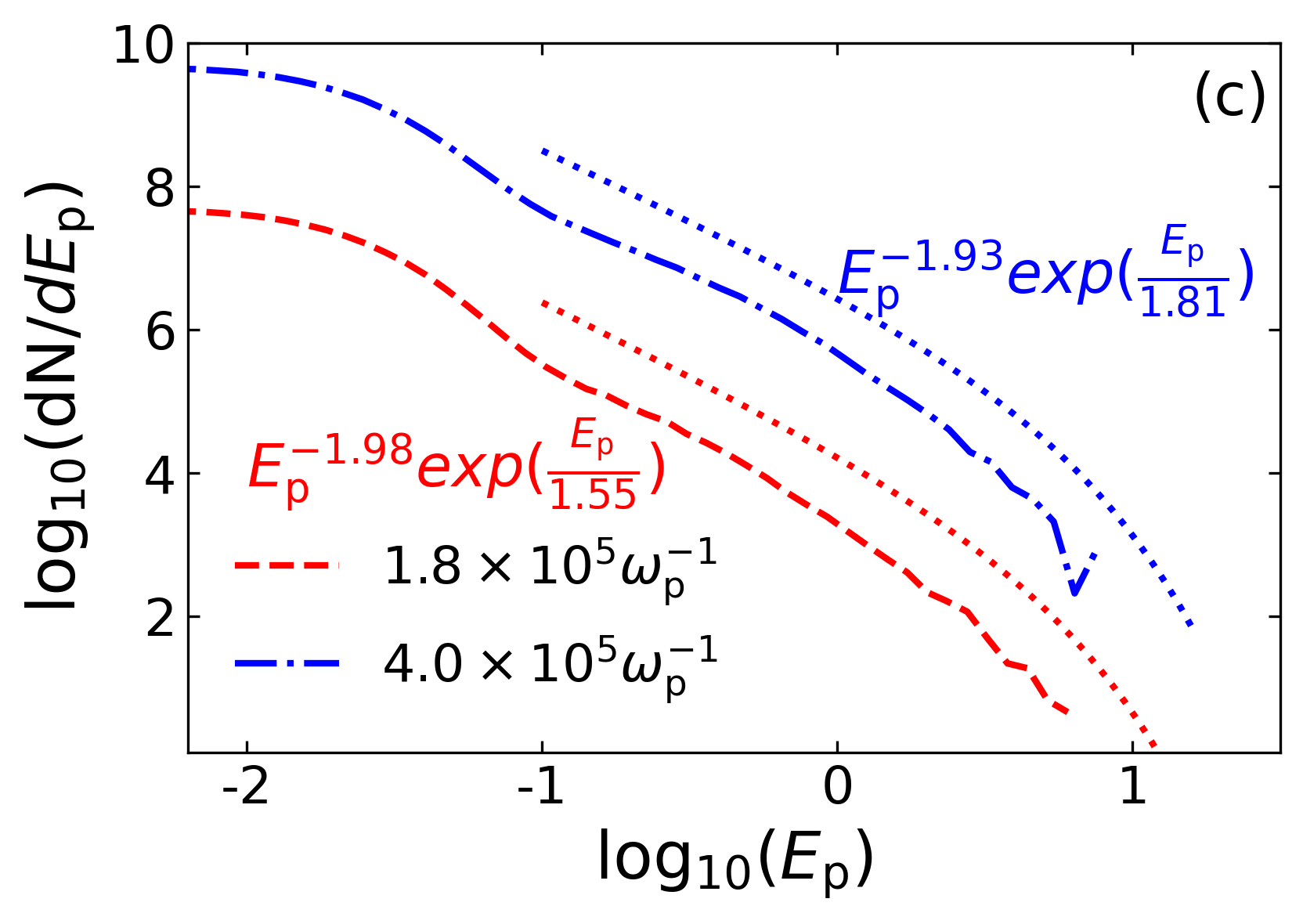}
    \caption{Panel (a): The average momentum (red line), parallel (blue dot-dashed line) and perpendicular (cyan dashed line) components
of all particles as a function of time, the data based on R-$\beta1$ case. 
    Panel (b): Same as panel (a) but for R-$\beta10$ case.
    panel (c): Particles energy spectra for $t=1.8\times10^5 \omega_{\rm p}^{-1}$ (red dashed line) and $t=4.0\times10^5 \omega_{\rm p}^{-1}$ (blue dot-dashed line).}
    \label{fig:mom_spectra}
\end{figure}
Given the lower effective $V_{\rm A}$ in high-$\beta$ relativistic cases, we extended the simulation for the R-$\beta10$ model to $4.0 \times10^5 \omega_{\rm p}^{-1}$ (approximately $10.5 t_{\rm A}$) to compare its evolution with that of the R-$\beta1$ model at $t=1.8 \times10^5 \omega_{\rm p}^{-1}$ ($\sim 10 t_{\rm A}$). Figure \ref{fig:mom_spectra} shows that the overall trends of particle momentum evolution are consistent between the two models, but the momentum growth rate in the R-$\beta10$ model is slower by about two orders of magnitude, consistent with the difference in cutoff energies listed in Table \ref{tab:parameters}. The observed trend is a direct consequence of how plasma $\beta$ regulates the Alfv\'en speed $V_{\rm A}$ and, consequently, the reconnection rate $V_{\rm rec}$ within the turbulent layer. In the high-$\beta$ regime, the increased thermal pressure (and relativistic enthalpy) raises the effective plasma inertia, reducing $V_{\rm A}$ and $V_{\rm rec}$. Therefore, the efficiency of the first-order Fermi acceleration within the turbulent reconnection layer also reduce (\citealt{deGouveiadalPino2005,Xu2023ApJ}), manifesting as a steeper power-law spectrum (larger $\alpha$, as shown Table \ref{tab:parameters} and Figure \ref{fig:mom_spectra}(c)). This physical explanation aligns with the scaling found in other studies of turbulent reconnection (e.g., \citealt{Takamoto2015ApJ,delvalle2016}).

In addition, we compared the particle energy spectra of the R-$\beta10$ model at $t=1.8 \times10^5 \omega_{\rm p}^{-1}$ and $t=4.0 \times10^5 \omega_{\rm p}^{-1}$ (Figure \ref{fig:mom_spectra}(c)). The fitting results show no significant change in either the spectral exponent $\alpha$ or the cutoff energy $E_{\rm cut}$. This indicates that in relativistic magnetic reconnection, although the global evolution is slower for high-$\beta$ models, the basic shape of the non-thermal particle energy spectrum is established at Fermi acceleration stage ($t\leq1.4\times 10^5\omega_{\rm p}^{-1}$. Extending the simulation time primarily leads to a broadening of the current sheet, which may involve more particles in the acceleration process and thus slightly hardens the spectrum (e.g., $\alpha$ adjusts from 1.98 to 1.93), but it does not alter the fundamental scaling relation of the spectral exponent. This further supports the statement that first-order Fermi acceleration determines the shape of the energy spectrum in the two-stage acceleration model (also see \citealt{Kowal2012,delvalle2016,Medina2021,Medina2023ApJ}).

\bibliography{sample701}{}

@ARTICLE{Bai2015,
       author = {{Bai}, Xue-Ning and {Caprioli}, Damiano and {Sironi}, Lorenzo and {Spitkovsky}, Anatoly},
        title = "{Magnetohydrodynamic-particle-in-cell Method for Coupling Cosmic Rays with a Thermal Plasma: Application to Non-relativistic Shocks}",
      journal = {\apj},
         year = 2015,
        month = aug,
       volume = {809},
       number = {1},
          eid = {55},
        pages = {55},
          doi = {10.1088/0004-637X/809/1/55},
archivePrefix = {arXiv},
       eprint = {1412.1087},
 primaryClass = {astro-ph.HE},
       adsurl = {https://ui.adsabs.harvard.edu/abs/2015ApJ...809...55B}
}

@ARTICLE{Ball2018ApJ,
       author = {{Ball}, David and {Sironi}, Lorenzo and {{\"O}zel}, Feryal},
        title = "{Electron and Proton Acceleration in Trans-relativistic Magnetic Reconnection: Dependence on Plasma Beta and Magnetization}",
      journal = {\apj},
         year = 2018,
        month = jul,
       volume = {862},
       number = {1},
          eid = {80},
        pages = {80},
          doi = {10.3847/1538-4357/aac820},
archivePrefix = {arXiv},
       eprint = {1803.05556},
 primaryClass = {astro-ph.HE},
       adsurl = {https://ui.adsabs.harvard.edu/abs/2018ApJ...862...80B}
}

@ARTICLE{Beresnyak2017,
       author = {{Beresnyak}, Andrey},
        title = "{Three-dimensional Spontaneous Magnetic Reconnection}",
      journal = {\apj},
         year = 2017,
        month = jan,
       volume = {834},
       number = {1},
          eid = {47},
        pages = {47},
          doi = {10.3847/1538-4357/834/1/47},
archivePrefix = {arXiv},
       eprint = {1301.7424},
 primaryClass = {astro-ph.SR},
       adsurl = {https://ui.adsabs.harvard.edu/abs/2017ApJ...834...47B}
}

@inproceedings{Boris1971,
  title = "{Proceedings of the Conference on the Numerical Simulation of Plasmas}",
  booktitle="{Proceedings of the Conference on the Numerical Simulation of Plasmas (4th) Held at the Naval Research Laboratory, Washington, D.C. on 2, 3 November 1970}",
  author={Jay P. Boris and Ramy A. Shanny},
  year = 1970,
  month = Nov,
}

@ARTICLE{Cao2021,
       author = {{Cao}, Zhen and {Aharonian}, F.~A. and {An}, Q. and {Axikegu}, L.~X., Bai and {Bai}, Y.~X. and {Bao}, Y.~W. and {Bastieri}, D. and {Bi}, X.~J. and {Bi}, Y.~J. and {Cai}, H. and {Cai}, J.~T. and {Cao}, Zhe and {Chang}, J. and {Chang}, J.~F. and {Chang}, X.~C. and {Chen}, B.~M. and {Chen}, J. and {Chen}, L. and {Chen}, Liang and {Chen}, Long and {Chen}, M.~J. and {Chen}, M.~L. and {Chen}, Q.~H. and {Chen}, S.~H. and {Chen}, S.~Z. and {Chen}, T.~L. and {Chen}, X.~L. and {Chen}, Y. and {Cheng}, N. and {Cheng}, Y.~D. and {Cui}, S.~W. and {Cui}, X.~H. and {Cui}, Y.~D. and {Dai}, B.~Z. and {Dai}, H.~L. and {Dai}, Z.~G. and {Danzengluobu} and {della Volpe}, D. and {D'Ettorre Piazzoli}, B. and {Dong}, X.~J. and {Fan}, J.~H. and {Fan}, Y.~Z. and {Fan}, Z.~X. and {Fang}, J. and {Fang}, K. and {Feng}, C.~F. and {Feng}, L. and {Feng}, S.~H. and {Feng}, Y.~L. and {Gao}, B. and {Gao}, C.~D. and {Gao}, Q. and {Gao}, W. and {Ge}, M.~M. and {Geng}, L.~S. and {Gong}, G.~H. and {Gou}, Q.~B. and {Gu}, M.~H. and {Guo}, J.~G. and {Guo}, X.~L. and {Guo}, Y.~Q. and {Guo}, Y.~Y. and {Han}, Y.~A. and {He}, H.~H. and {He}, H.~N. and {He}, J.~C. and {He}, S.~L. and {He}, X.~B. and {He}, Y. and {Heller}, M. and {Hor}, Y.~K. and {Hou}, C. and {Hou}, X. and {Hu}, H.~B. and {Hu}, S. and {Hu}, S.~C. and {Hu}, X.~J. and {Huang}, D.~H. and {Huang}, Q.~L. and {Huang}, W.~H. and {Huang}, X.~T. and {Huang}, Z.~C. and {Ji}, F. and {Ji}, X.~L. and {Jia}, H.~Y. and {Jiang}, K. and {Jiang}, Z.~J. and {Jin}, C. and {Kuleshov}, D. and {Levochkin}, K. and {Li}, B.~B. and {Li}, Cong and {Li}, Cheng and {Li}, F. and {Li}, H.~B. and {Li}, H.~C. and {Li}, H.~Y. and {Li}, J. and {Li}, K. and {Li}, W.~L. and {Li}, X. and {Li}, Xin and {Li}, X.~R. and {Li}, Y. and {Li}, Y.~Z. and {Li}, Zhe and {Li}, Zhuo and {Liang}, E.~W. and {Liang}, Y.~F. and {Lin}, S.~J. and {Liu}, B. and {Liu}, C. and {Liu}, D. and {Liu}, H. and {Liu}, H.~D. and {Liu}, J. and {Liu}, J.~L. and {Liu}, J.~S. and {Liu}, J.~Y. and {Liu}, M.~Y. and {Liu}, R.~Y. and {Liu}, S.~M. and {Liu}, W. and {Liu}, Y.~N. and {Liu}, Z.~X. and {Long}, W.~J. and {Lu}, R. and {Lv}, H.~K. and {Ma}, B.~Q. and {Ma}, L.~L. and {Ma}, X.~H. and {Mao}, J.~R. and {Masood}, A. and {Mitthumsiri}, W. and {Montaruli}, T. and {Nan}, Y.~C. and {Pang}, B.~Y. and {Pattarakijwanich}, P. and {Pei}, Z.~Y. and {Qi}, M.~Y. and {Ruffolo}, D. and {Rulev}, V. and {S{\'a}iz}, A. and {Shao}, L. and {Shchegolev}, O. and {Sheng}, X.~D. and {Shi}, J.~R. and {Song}, H.~C. and {Stenkin}, Yu. V. and {Stepanov}, V. and {Sun}, Q.~N. and {Sun}, X.~N. and {Sun}, Z.~B. and {Tam}, P.~H.~T. and {Tang}, Z.~B. and {Tian}, W.~W. and {Wang}, B.~D. and {Wang}, C. and {Wang}, H. and {Wang}, H.~G. and {Wang}, J.~C. and {Wang}, J.~S. and {Wang}, L.~P. and {Wang}, L.~Y. and {Wang}, R.~N. and {Wang}, W. and {Wang}, W. and {Wang}, X.~G. and {Wang}, X.~J. and {Wang}, X.~Y. and {Wang}, Y.~D. and {Wang}, Y.~J. and {Wang}, Y.~P. and {Wang}, Zheng and {Wang}, Zhen and {Wang}, Z.~H. and {Wang}, Z.~X. and {Wei}, D.~M. and {Wei}, J.~J. and {Wei}, Y.~J. and {Wen}, T. and {Wu}, C.~Y. and {Wu}, H.~R. and {Wu}, S. and {Wu}, W.~X. and {Wu}, X.~F. and {Xi}, S.~Q. and {Xia}, J. and {Xia}, J.~J. and {Xiang}, G.~M. and {Xiao}, G. and {Xiao}, H.~B. and {Xin}, G.~G. and {Xin}, Y.~L. and {Xing}, Y. and {Xu}, D.~L. and {Xu}, R.~X. and {Xue}, L. and {Yan}, D.~H. and {Yang}, C.~W.},
        title = "{Ultrahigh-energy photons up to 1.4 petaelectronvolts from 12 {\ensuremath{\gamma}}-ray Galactic sources}",
      journal = {\nat},
         year = 2021,
        month = jun,
       volume = {594},
       number = {7861},
        pages = {33-36},
          doi = {10.1038/s41586-021-03498-z},
       adsurl = {https://ui.adsabs.harvard.edu/abs/2021Natur.594...33C}
}

@ARTICLE{Cerutti2014b,
       author = {{Cerutti}, B. and {Werner}, G.~R. and {Uzdensky}, D.~A. and {Begelman}, M.~C.},
        title = "{Three-dimensional Relativistic Pair Plasma Reconnection with Radiative Feedback in the Crab Nebula}",
      journal = {\apj},
         year = 2014,
        month = feb,
       volume = {782},
       number = {2},
          eid = {104},
        pages = {104},
          doi = {10.1088/0004-637X/782/2/104},
archivePrefix = {arXiv},
       eprint = {1311.2605},
 primaryClass = {astro-ph.HE},
       adsurl = {https://ui.adsabs.harvard.edu/abs/2014ApJ...782..104C}
}

@ARTICLE{Christie2019MNRAS,
       author = {{Christie}, I.~M. and {Petropoulou}, M. and {Sironi}, L. and {Giannios}, D.},
        title = "{Radiative signatures of plasmoid-dominated reconnection in blazar jets}",
      journal = {\mnras},
         year = 2019,
        month = jan,
       volume = {482},
       number = {1},
        pages = {65-82},
          doi = {10.1093/mnras/sty2636},
archivePrefix = {arXiv},
       eprint = {1807.08041},
 primaryClass = {astro-ph.HE},
       adsurl = {https://ui.adsabs.harvard.edu/abs/2019MNRAS.482...65C}
}

@ARTICLE{deGouveiadalPino2005,
       author = {{de Gouveia dal Pino}, E.~M. and {Lazarian}, A.},
        title = "{Production of the large scale superluminal ejections of the microquasar GRS 1915+105 by violent magnetic reconnection}",
      journal = {\aap},
         year = 2005,
        month = oct,
       volume = {441},
       number = {3},
        pages = {845-853},
          doi = {10.1051/0004-6361:20042590},
       adsurl = {https://ui.adsabs.harvard.edu/abs/2005A&A...441..845D}
}

@ARTICLE{deGouveiadalPino2024arXiv,
       author = {{de Gouveia Dal Pino}, Elisabete M. and {Medina-Torrejon}, Tania E.},
        title = "{Particle Acceleration Time due to Turbulent-Induced Magnetic Reconnection}",
      journal = {arXiv e-prints},
         year = 2024,
        month = oct,
          eid = {arXiv:2410.13071},
        pages = {arXiv:2410.13071},
          doi = {10.48550/arXiv.2410.13071},
archivePrefix = {arXiv},
       eprint = {2410.13071},
 primaryClass = {astro-ph.HE},
       adsurl = {https://ui.adsabs.harvard.edu/abs/2024arXiv241013071D}
}

@ARTICLE{deGouveiaDalPino2025MNRAS,
       author = {{de Gouveia Dal Pino}, E.~M. and {Rodr{\'\i}guez-Ram{\'\i}rez}, J.~C. and {del Valle}, M.~V.},
        title = "{Multimessenger emission from magnetic reconnection in blazar jets: the case of TXS 0506+056}",
      journal = {\mnras},
         year = 2025,
        month = mar,
       volume = {537},
       number = {4},
        pages = {3895-3907},
          doi = {10.1093/mnras/staf251},
archivePrefix = {arXiv},
       eprint = {2411.10210},
 primaryClass = {astro-ph.HE},
       adsurl = {https://ui.adsabs.harvard.edu/abs/2025MNRAS.537.3895D}
}

@ARTICLE{delvalle2016,
       author = {{del Valle}, Maria V. and {de Gouveia Dal Pino}, E.~M. and {Kowal}, G.},
        title = "{Properties of the first-order Fermi acceleration in fast magnetic reconnection driven by turbulence in collisional magnetohydrodynamical flows}",
      journal = {\mnras},
         year = 2016,
        month = dec,
       volume = {463},
       number = {4},
        pages = {4331-4343},
          doi = {10.1093/mnras/stw2276},
archivePrefix = {arXiv},
       eprint = {1609.08598},
 primaryClass = {astro-ph.HE},
       adsurl = {https://ui.adsabs.harvard.edu/abs/2016MNRAS.463.4331D}
}

@ARTICLE{Drake2010,
       author = {{Drake}, J.~F. and {Opher}, M. and {Swisdak}, M. and {Chamoun}, J.~N.},
        title = "{A Magnetic Reconnection Mechanism for the Generation of Anomalous Cosmic Rays}",
      journal = {\apj},
         year = 2010,
        month = feb,
       volume = {709},
       number = {2},
        pages = {963-974},
          doi = {10.1088/0004-637X/709/2/963},
archivePrefix = {arXiv},
       eprint = {0911.3098},
 primaryClass = {astro-ph.SR},
       adsurl = {https://ui.adsabs.harvard.edu/abs/2010ApJ...709..963D}
}

@ARTICLE{Eyink2011,
       author = {{Eyink}, Gregory L. and {Lazarian}, A. and {Vishniac}, E.~T.},
        title = "{Fast Magnetic Reconnection and Spontaneous Stochasticity}",
      journal = {\apj},
         year = 2011,
        month = dec,
       volume = {743},
       number = {1},
          eid = {51},
        pages = {51},
          doi = {10.1088/0004-637X/743/1/51},
archivePrefix = {arXiv},
       eprint = {1103.1882},
 primaryClass = {astro-ph.GA},
       adsurl = {https://ui.adsabs.harvard.edu/abs/2011ApJ...743...51E}
}

@ARTICLE{Eyink2013,
       author = {{Eyink}, Gregory and {Vishniac}, Ethan and {Lalescu}, Cristian and {Aluie}, Hussein and {Kanov}, Kalin and {B{\"u}rger}, Kai and {Burns}, Randal and {Meneveau}, Charles and {Szalay}, Alexander},
        title = "{Flux-freezing breakdown in high-conductivity magnetohydrodynamic turbulence}",
      journal = {\nat},
         year = 2013,
        month = may,
       volume = {497},
       number = {7450},
        pages = {466-469},
          doi = {10.1038/nature12128},
       adsurl = {https://ui.adsabs.harvard.edu/abs/2013Natur.497..466E}
}

@ARTICLE{Fabian2015MNRAS,
       author = {{Fabian}, A.~C. and {Lohfink}, A. and {Kara}, E. and {Parker}, M.~L. and {Vasudevan}, R. and {Reynolds}, C.~S.},
        title = "{Properties of AGN coronae in the NuSTAR era}",
      journal = {\mnras},
         year = 2015,
        month = aug,
       volume = {451},
       number = {4},
        pages = {4375-4383},
          doi = {10.1093/mnras/stv1218},
archivePrefix = {arXiv},
       eprint = {1505.07603},
 primaryClass = {astro-ph.HE},
       adsurl = {https://ui.adsabs.harvard.edu/abs/2015MNRAS.451.4375F}
}

@ARTICLE{Guo2014,
       author = {{Guo}, Fan and {Li}, Hui and {Daughton}, William and {Liu}, Yi-Hsin},
        title = "{Formation of Hard Power Laws in the Energetic Particle Spectra Resulting from Relativistic Magnetic Reconnection}",
      journal = {\prl},
         year = 2014,
        month = oct,
       volume = {113},
       number = {15},
          eid = {155005},
        pages = {155005},
          doi = {10.1103/PhysRevLett.113.155005},
archivePrefix = {arXiv},
       eprint = {1405.4040},
 primaryClass = {astro-ph.HE},
       adsurl = {https://ui.adsabs.harvard.edu/abs/2014PhRvL.113o5005G}
}

@ARTICLE{Guo2015,
       author = {{Guo}, Fan and {Liu}, Yi-Hsin and {Daughton}, William and {Li}, Hui},
        title = "{Particle Acceleration and Plasma Dynamics during Magnetic Reconnection in the Magnetically Dominated Regime}",
      journal = {\apj},
         year = 2015,
        month = jun,
       volume = {806},
       number = {2},
          eid = {167},
        pages = {167},
          doi = {10.1088/0004-637X/806/2/167},
archivePrefix = {arXiv},
       eprint = {1504.02193},
 primaryClass = {astro-ph.HE},
       adsurl = {https://ui.adsabs.harvard.edu/abs/2015ApJ...806..167G}
}

@ARTICLE{Guo2021,
       author = {{Guo}, Fan and {Li}, Xiaocan and {Daughton}, William and {Li}, Hui and {Kilian}, Patrick and {Liu}, Yi-Hsin and {Zhang}, Qile and {Zhang}, Haocheng},
        title = "{Magnetic Energy Release, Plasma Dynamics, and Particle Acceleration in Relativistic Turbulent Magnetic Reconnection}",
      journal = {\apj},
         year = 2021,
        month = oct,
       volume = {919},
       number = {2},
          eid = {111},
        pages = {111},
          doi = {10.3847/1538-4357/ac0918},
archivePrefix = {arXiv},
       eprint = {2008.02743},
 primaryClass = {astro-ph.HE},
       adsurl = {https://ui.adsabs.harvard.edu/abs/2021ApJ...919..111G}
}

@ARTICLE{Harris1962,
       author = {{Harris}, E.~G.},
        title = "{On a plasma sheath separating regions of oppositely directed magnetic field}",
      journal = {Il Nuovo Cimento},
         year = 1962,
        month = jan,
       volume = {23},
       number = {1},
        pages = {115-121},
          doi = {10.1007/BF02733547},
       adsurl = {https://ui.adsabs.harvard.edu/abs/1962NCim...23..115H}
}

@ARTICLE{Kadowaki2015,
       author = {{Kadowaki}, L.~H.~S. and {de Gouveia Dal Pino}, E.~M. and {Singh}, C.~B.},
        title = "{The Role of Fast Magnetic Reconnection on the Radio and Gamma-ray Emission from the Nuclear Regions of Microquasars and Low Luminosity AGNs}",
      journal = {\apj},
         year = 2015,
        month = apr,
       volume = {802},
       number = {2},
          eid = {113},
        pages = {113},
          doi = {10.1088/0004-637X/802/2/113},
archivePrefix = {arXiv},
       eprint = {1410.3454},
 primaryClass = {astro-ph.HE},
       adsurl = {https://ui.adsabs.harvard.edu/abs/2015ApJ...802..113K}
}

@ARTICLE{Kadowaki2018ApJ,
       author = {{Kadowaki}, Luis H.~S. and {De Gouveia Dal Pino}, Elisabete M. and {Stone}, James M.},
        title = "{MHD Instabilities in Accretion Disks and Their Implications in Driving Fast Magnetic Reconnection}",
      journal = {\apj},
         year = 2018,
        month = sep,
       volume = {864},
       number = {1},
          eid = {52},
        pages = {52},
          doi = {10.3847/1538-4357/aad4ff},
archivePrefix = {arXiv},
       eprint = {1803.08557},
 primaryClass = {astro-ph.HE},
       adsurl = {https://ui.adsabs.harvard.edu/abs/2018ApJ...864...52K}
}

@ARTICLE{Kadowaki2021,
       author = {{Kadowaki}, Luis H.~S. and {de Gouveia Dal Pino}, Elisabete M. and {Medina-Torrej{\'o}n}, Tania E. and {Mizuno}, Yosuke and {Kushwaha}, Pankaj},
        title = "{Fast Magnetic Reconnection Structures in Poynting Flux-dominated Jets}",
      journal = {\apj},
         year = 2021,
        month = may,
       volume = {912},
       number = {2},
          eid = {109},
        pages = {109},
          doi = {10.3847/1538-4357/abee7a},
archivePrefix = {arXiv},
       eprint = {2011.03634},
 primaryClass = {astro-ph.HE},
       adsurl = {https://ui.adsabs.harvard.edu/abs/2021ApJ...912..109K}
}

@ARTICLE{Kowal2009,
       author = {{Kowal}, Grzegorz and {Lazarian}, A. and {Vishniac}, E.~T. and {Otmianowska-Mazur}, K.},
        title = "{Numerical Tests of Fast Reconnection in Weakly Stochastic Magnetic Fields}",
      journal = {\apj},
         year = 2009,
        month = jul,
       volume = {700},
       number = {1},
        pages = {63-85},
          doi = {10.1088/0004-637X/700/1/63},
archivePrefix = {arXiv},
       eprint = {0903.2052},
 primaryClass = {astro-ph.GA},
       adsurl = {https://ui.adsabs.harvard.edu/abs/2009ApJ...700...63K}
}

@ARTICLE{Kowal2011,
       author = {{Kowal}, Grzegorz and {de Gouveia Dal Pino}, E.~M. and {Lazarian}, A.},
        title = "{Magnetohydrodynamic Simulations of Reconnection and Particle Acceleration: Three-dimensional Effects}",
      journal = {\apj},
         year = 2011,
        month = jul,
       volume = {735},
       number = {2},
          eid = {102},
        pages = {102},
          doi = {10.1088/0004-637X/735/2/102},
archivePrefix = {arXiv},
       eprint = {1103.2984},
 primaryClass = {astro-ph.HE},
       adsurl = {https://ui.adsabs.harvard.edu/abs/2011ApJ...735..102K}
}

@ARTICLE{Kowal2012,
       author = {{Kowal}, Grzegorz and {de Gouveia Dal Pino}, Elisabete M. and {Lazarian}, A.},
        title = "{Particle Acceleration in Turbulence and Weakly Stochastic Reconnection}",
      journal = {\prl},
         year = 2012,
        month = jun,
       volume = {108},
       number = {24},
          eid = {241102},
        pages = {241102},
          doi = {10.1103/PhysRevLett.108.241102},
archivePrefix = {arXiv},
       eprint = {1202.5256},
 primaryClass = {astro-ph.HE},
       adsurl = {https://ui.adsabs.harvard.edu/abs/2012PhRvL.108x1102K}
}

@ARTICLE{Kowal2020,
       author = {{Kowal}, Grzegorz and {Falceta-Gon{\c{c}}alves}, Diego A. and {Lazarian}, Alex and {Vishniac}, Ethan T.},
        title = "{Kelvin-Helmholtz versus Tearing Instability: What Drives Turbulence in Stochastic Reconnection?}",
      journal = {\apj},
         year = 2020,
        month = mar,
       volume = {892},
       number = {1},
          eid = {50},
        pages = {50},
          doi = {10.3847/1538-4357/ab7a13},
archivePrefix = {arXiv},
       eprint = {1909.09179},
 primaryClass = {astro-ph.HE},
       adsurl = {https://ui.adsabs.harvard.edu/abs/2020ApJ...892...50K}
}

@ARTICLE{Kumar2015,
       author = {{Kumar}, Pawan and {Zhang}, Bing},
        title = "{The physics of gamma-ray bursts \& relativistic jets}",
      journal = {\physrep},
         year = 2015,
        month = feb,
       volume = {561},
        pages = {1-109},
          doi = {10.1016/j.physrep.2014.09.008},
archivePrefix = {arXiv},
       eprint = {1410.0679},
 primaryClass = {astro-ph.HE},
       adsurl = {https://ui.adsabs.harvard.edu/abs/2015PhR...561....1K}
}

@article{Landi2015,
doi = {10.1088/0004-637X/806/1/131},
url = {https://dx.doi.org/10.1088/0004-637X/806/1/131},
year = {2015},
month = {jun},
publisher = {The American Astronomical Society},
volume = {806},
number = {1},
pages = {131},
author = {S. Landi and L. Del Zanna and E. Papini and F. Pucci and M. Velli},
title = {RESISTIVE MAGNETOHYDRODYNAMICS SIMULATIONS OF THE IDEAL TEARING MODE},
journal = {The Astrophysical Journal},
}

@ARTICLE{Lazarian1999,
       author = {{Lazarian}, A. and {Vishniac}, Ethan T.},
        title = "{Reconnection in a Weakly Stochastic Field}",
      journal = {\apj},
         year = 1999,
        month = jun,
       volume = {517},
       number = {2},
        pages = {700-718},
          doi = {10.1086/307233},
archivePrefix = {arXiv},
       eprint = {astro-ph/9811037},
 primaryClass = {astro-ph},
       adsurl = {https://ui.adsabs.harvard.edu/abs/1999ApJ...517..700L}
}

@ARTICLE{Lazarian2020,
       author = {{Lazarian}, Alex and {Eyink}, Gregory L. and {Jafari}, Amir and {Kowal}, Grzegorz and {Li}, Hui and {Xu}, Siyao and {Vishniac}, Ethan T.},
        title = "{3D turbulent reconnection: Theory, tests, and astrophysical implications}",
      journal = {Physics of Plasmas},
         year = 2020,
        month = jan,
       volume = {27},
       number = {1},
          eid = {012305},
        pages = {012305},
          doi = {10.1063/1.5110603},
archivePrefix = {arXiv},
       eprint = {2001.00868},
 primaryClass = {astro-ph.HE},
       adsurl = {https://ui.adsabs.harvard.edu/abs/2020PhPl...27a2305L}
}

@ARTICLE{Liang2023,
       author = {{Liang}, Shi-Min and {Zhang}, Jian-Fu and {Gao}, Na-Na and {Xiao}, Hua-Ping},
        title = "{Magnetic-reconnection-driven Turbulence and Turbulent Reconnection Acceleration}",
      journal = {\apj},
         year = 2023,
        month = aug,
       volume = {952},
       number = {2},
          eid = {93},
        pages = {93},
          doi = {10.3847/1538-4357/acdc18},
archivePrefix = {arXiv},
       eprint = {2306.03418},
 primaryClass = {astro-ph.HE},
       adsurl = {https://ui.adsabs.harvard.edu/abs/2023ApJ...952...93L}
}

@ARTICLE{Liang2025AA,
       author = {{Liang}, Shi-Min and {Zhang}, Jian-Fu and {Gao}, Na-Na and {Yi}, Nian-Yu},
        title = "{Studying the properties of reconnection-driven turbulence}",
      journal = {\aap},
         year = 2025,
        month = nov,
       volume = {703},
          eid = {A226},
        pages = {A226},
          doi = {10.1051/0004-6361/202553812},
archivePrefix = {arXiv},
       eprint = {2510.09978},
 primaryClass = {astro-ph.HE},
       adsurl = {https://ui.adsabs.harvard.edu/abs/2025A&A...703A.226L}
}

@ARTICLE{Liang2025PoP,
       author = {{Liang}, Shimin and {Yi}, Nianyu},
        title = "{The role of particle feedback on particle acceleration in magnetic reconnection}",
      journal = {arXiv e-prints},
         year = 2025,
        month = dec,
          eid = {arXiv:2512.24054},
        pages = {arXiv:2512.24054},
          doi = {10.48550/arXiv.2512.24054},
archivePrefix = {arXiv},
       eprint = {2512.24054},
 primaryClass = {physics.plasm-ph},
       adsurl = {https://ui.adsabs.harvard.edu/abs/2025arXiv251224054L}
}

@ARTICLE{Liu2009,
       author = {{Liu}, W.~J. and {Chen}, P.~F. and {Ding}, M.~D. and {Fang}, C.},
        title = "{Energy Spectrum of the Electrons Accelerated by a Reconnection Electric Field: Exponential or Power Law?}",
      journal = {\apj},
         year = 2009,
        month = jan,
       volume = {690},
       number = {2},
        pages = {1633-1638},
          doi = {10.1088/0004-637X/690/2/1633},
archivePrefix = {arXiv},
       eprint = {0809.1212},
 primaryClass = {astro-ph},
       adsurl = {https://ui.adsabs.harvard.edu/abs/2009ApJ...690.1633L}
}

@ARTICLE{Medina2021,
       author = {{Medina-Torrej{\'o}n}, Tania E. and {de Gouveia Dal Pino}, Elisabete M. and {Kadowaki}, Luis H.~S. and {Kowal}, Grzegorz and {Singh}, Chandra B. and {Mizuno}, Yosuke},
        title = "{Particle Acceleration by Relativistic Magnetic Reconnection Driven by Kink Instability Turbulence in Poynting Flux-Dominated Jets}",
      journal = {\apj},
         year = 2021,
        month = feb,
       volume = {908},
       number = {2},
          eid = {193},
        pages = {193},
          doi = {10.3847/1538-4357/abd6c2},
archivePrefix = {arXiv},
       eprint = {2009.08516},
 primaryClass = {astro-ph.HE},
       adsurl = {https://ui.adsabs.harvard.edu/abs/2021ApJ...908..193M}
}

@ARTICLE{Medina2023ApJ,
       author = {{Medina-Torrej{\'o}n}, Tania E. and {de Gouveia Dal Pino}, Elisabete M. and {Kowal}, Grzegorz},
        title = "{Particle Acceleration by Magnetic Reconnection in Relativistic Jets: The Transition from Small to Large Scales}",
      journal = {\apj},
         year = 2023,
        month = aug,
       volume = {952},
       number = {2},
          eid = {168},
        pages = {168},
          doi = {10.3847/1538-4357/acd699},
archivePrefix = {arXiv},
       eprint = {2303.08780},
 primaryClass = {astro-ph.HE},
       adsurl = {https://ui.adsabs.harvard.edu/abs/2023ApJ...952..168M}
}

@ARTICLE{Mignone2007,
       author = {{Mignone}, A. and {Bodo}, G. and {Massaglia}, S. and {Matsakos}, T. and {Tesileanu}, O. and {Zanni}, C. and {Ferrari}, A.},
        title = "{PLUTO: A Numerical Code for Computational Astrophysics}",
      journal = {\apjs},
         year = 2007,
        month = may,
       volume = {170},
       number = {1},
        pages = {228-242},
          doi = {10.1086/513316},
archivePrefix = {arXiv},
       eprint = {astro-ph/0701854},
 primaryClass = {astro-ph},
       adsurl = {https://ui.adsabs.harvard.edu/abs/2007ApJS..170..228M}
}

@ARTICLE{Mignone2018,
       author = {{Mignone}, A. and {Bodo}, G. and {Vaidya}, B. and {Mattia}, G.},
        title = "{A Particle Module for the PLUTO Code. I. An Implementation of the MHD-PIC Equations}",
      journal = {\apj},
         year = 2018,
        month = may,
       volume = {859},
       number = {1},
          eid = {13},
        pages = {13},
          doi = {10.3847/1538-4357/aabccd},
archivePrefix = {arXiv},
       eprint = {1804.01946},
 primaryClass = {astro-ph.HE},
       adsurl = {https://ui.adsabs.harvard.edu/abs/2018ApJ...859...13M}
}

@INPROCEEDINGS{Mignone2020,
       author = {{Mignone}, A. and {Vaidya}, B. and {Puzzoni}, E. and {Mukherjee}, D. and {Bodo}, G. and {Flock}, M.},
        title = "{Particle-Gas Hybrid Schemes in the PLUTO Code}",
    booktitle = {Journal of Physics Conference Series},
         year = 2020,
       series = {Journal of Physics Conference Series},
       volume = {1623},
        month = sep,
          eid = {012007},
        pages = {012007},
          doi = {10.1088/1742-6596/1623/1/012007},
       adsurl = {https://ui.adsabs.harvard.edu/abs/2020JPhCS1623a2007M}
}

@ARTICLE{Mora2025,
       author = {{Mora}, Clarissa and {Bacchini}, Fabio and {Keppens}, Rony},
        title = "{Nonthermal electron acceleration in turbulent post-flare coronal loops}",
      journal = {arXiv e-prints},
         year = 2025,
        month = oct,
          eid = {arXiv:2510.18742},
        pages = {arXiv:2510.18742},
          doi = {10.48550/arXiv.2510.18742},
archivePrefix = {arXiv},
       eprint = {2510.18742},
 primaryClass = {astro-ph.SR},
       adsurl = {https://ui.adsabs.harvard.edu/abs/2025arXiv251018742M}
}

@ARTICLE{Nava2011AA,
       author = {{Nava}, L. and {Ghirlanda}, G. and {Ghisellini}, G. and {Celotti}, A.},
        title = "{Spectral properties of 438 GRBs detected by Fermi/GBM}",
      journal = {\aap},
         year = 2011,
        month = jun,
       volume = {530},
          eid = {A21},
        pages = {A21},
          doi = {10.1051/0004-6361/201016270},
archivePrefix = {arXiv},
       eprint = {1012.2863},
 primaryClass = {astro-ph.HE},
       adsurl = {https://ui.adsabs.harvard.edu/abs/2011A&A...530A..21N}
}

@ARTICLE{Parker1957,
       author = {{Parker}, E.~N.},
        title = "{Sweet's Mechanism for Merging Magnetic Fields in Conducting Fluids}",
      journal = {\jgr},
         year = 1957,
        month = dec,
       volume = {62},
       number = {4},
        pages = {509-520},
          doi = {10.1029/JZ062i004p00509},
       adsurl = {https://ui.adsabs.harvard.edu/abs/1957JGR....62..509P}
}

@article{Papini2019a,
          doi = {10.3847/1538-4357/ab4352},
          url = {https://dx.doi.org/10.3847/1538-4357/ab4352},
         year = {2019},
        month = {oct},
    publisher = {The American Astronomical Society},
       volume = {885},
       number = {1},
        pages = {56},
       author = {E. Papini and S. Landi and L. Del Zanna},
        title = {Fast Magnetic Reconnection: Secondary Tearing Instability and Role of the Hall Term},
      journal = {The Astrophysical Journal},
}

@ARTICLE{Remillard2006ARAA,
       author = {{Remillard}, Ronald A. and {McClintock}, Jeffrey E.},
        title = "{X-Ray Properties of Black-Hole Binaries}",
      journal = {\araa},
         year = 2006,
        month = sep,
       volume = {44},
       number = {1},
        pages = {49-92},
          doi = {10.1146/annurev.astro.44.051905.092532},
archivePrefix = {arXiv},
       eprint = {astro-ph/0606352},
 primaryClass = {astro-ph},
       adsurl = {https://ui.adsabs.harvard.edu/abs/2006ARA&A..44...49R}
}

@ARTICLE{Ripperda2017a,
       author = {{Ripperda}, B. and {Porth}, O. and {Xia}, C. and {Keppens}, R.},
        title = "{Reconnection and particle acceleration in interacting flux ropes - I. Magnetohydrodynamics and test particles in 2.5D}",
      journal = {\mnras},
         year = 2017,
        month = may,
       volume = {467},
       number = {3},
        pages = {3279-3298},
          doi = {10.1093/mnras/stx379},
archivePrefix = {arXiv},
       eprint = {1611.09966},
 primaryClass = {astro-ph.HE},
       adsurl = {https://ui.adsabs.harvard.edu/abs/2017MNRAS.467.3279R}
}

@ARTICLE{Rodriguez-Ramirez2019ApJ,
       author = {{Rodr{\'\i}guez-Ram{\'\i}rez}, Juan Carlos and {de Gouveia Dal Pino}, Elisabete M. and {Alves Batista}, Rafael},
        title = "{Very-high-energy Emission from Magnetic Reconnection in the Radiative-inefficient Accretion Flow of SgrA*}",
      journal = {\apj},
         year = 2019,
        month = jul,
       volume = {879},
       number = {1},
          eid = {6},
        pages = {6},
          doi = {10.3847/1538-4357/ab212e},
archivePrefix = {arXiv},
       eprint = {1904.05765},
 primaryClass = {astro-ph.HE},
       adsurl = {https://ui.adsabs.harvard.edu/abs/2019ApJ...879....6R}
}

@ARTICLE{Singh2015ApJ,
       author = {{Singh}, C.~B. and {de Gouveia Dal Pino}, E.~M. and {Kadowaki}, L.~H.~S.},
        title = "{On the Role of Fast Magnetic Reconnection in Accreting Black Hole Sources}",
      journal = {\apjl},
         year = 2015,
        month = feb,
       volume = {799},
       number = {2},
          eid = {L20},
        pages = {L20},
          doi = {10.1088/2041-8205/799/2/L20},
archivePrefix = {arXiv},
       eprint = {1411.0883},
 primaryClass = {astro-ph.HE},
       adsurl = {https://ui.adsabs.harvard.edu/abs/2015ApJ...799L..20S}
}

@ARTICLE{Sironi2014,
       author = {{Sironi}, Lorenzo and {Spitkovsky}, Anatoly},
        title = "{Relativistic Reconnection: An Efficient Source of Non-thermal Particles}",
      journal = {\apjl},
         year = 2014,
        month = mar,
       volume = {783},
       number = {1},
          eid = {L21},
        pages = {L21},
          doi = {10.1088/2041-8205/783/1/L21},
archivePrefix = {arXiv},
       eprint = {1401.5471},
 primaryClass = {astro-ph.HE},
       adsurl = {https://ui.adsabs.harvard.edu/abs/2014ApJ...783L..21S}
}

@ARTICLE{Sironi2015MNRAS,
       author = {{Sironi}, Lorenzo and {Petropoulou}, Maria and {Giannios}, Dimitrios},
        title = "{Relativistic jets shine through shocks or magnetic reconnection?}",
      journal = {\mnras},
         year = 2015,
        month = jun,
       volume = {450},
       number = {1},
        pages = {183-191},
          doi = {10.1093/mnras/stv641},
archivePrefix = {arXiv},
       eprint = {1502.01021},
 primaryClass = {astro-ph.HE},
       adsurl = {https://ui.adsabs.harvard.edu/abs/2015MNRAS.450..183S}
}

@ARTICLE{Sweet1958,
       author = {{Sweet}, P.~A.},
        title = "{The topology of force-free magnetic fields}",
      journal = {The Observatory},
         year = 1958,
        month = feb,
       volume = {78},
        pages = {30-32},
       adsurl = {https://ui.adsabs.harvard.edu/abs/1958Obs....78...30S}
}

@ARTICLE{Takamoto2015ApJ,
       author = {{Takamoto}, Makoto and {Inoue}, Tsuyoshi and {Lazarian}, Alexandre},
        title = "{Turbulent Reconnection in Relativistic Plasmas and Effects of Compressibility}",
      journal = {\apj},
         year = 2015,
        month = dec,
       volume = {815},
       number = {1},
          eid = {16},
        pages = {16},
          doi = {10.1088/0004-637X/815/1/16},
archivePrefix = {arXiv},
       eprint = {1509.07703},
 primaryClass = {astro-ph.HE},
       adsurl = {https://ui.adsabs.harvard.edu/abs/2015ApJ...815...16T}
}

@ARTICLE{Uzdensky2014,
       author = {{Uzdensky}, Dmitri A. and {Spitkovsky}, Anatoly},
        title = "{Physical Conditions in the Reconnection Layer in Pulsar Magnetospheres}",
      journal = {\apj},
         year = 2014,
        month = jan,
       volume = {780},
       number = {1},
          eid = {3},
        pages = {3},
          doi = {10.1088/0004-637X/780/1/3},
archivePrefix = {arXiv},
       eprint = {1210.3346},
 primaryClass = {astro-ph.HE},
       adsurl = {https://ui.adsabs.harvard.edu/abs/2014ApJ...780....3U}
}

@ARTICLE{Vicentin2025,
       author = {{Vicentin}, Giovani H. and {Kowal}, Grzegorz and {Dal Pino}, Elisabete M. de Gouveia and {Lazarian}, Alex},
        title = "{Investigating Turbulence Effects on Magnetic Reconnection Rates through 3D Resistive Magnetohydrodynamic Simulations}",
      journal = {\apj},
         year = 2025,
        month = jul,
       volume = {987},
       number = {2},
          eid = {213},
        pages = {213},
          doi = {10.3847/1538-4357/addc62},
archivePrefix = {arXiv},
       eprint = {2405.15909},
 primaryClass = {physics.plasm-ph},
       adsurl = {https://ui.adsabs.harvard.edu/abs/2025ApJ...987..213V}
}

@ARTICLE{Vicentin2025arXiv,
       author = {{Vicentin}, G.~H. and {Kowal}, G. and {de Gouveia Dal Pino}, E.~M. and {Lazarian}, A.},
        title = "{Do plasmoids induce fast magnetic reconnection in well-resolved current sheets in 2D MHD simulations?}",
      journal = {arXiv e-prints},
         year = 2025,
        month = oct,
          eid = {arXiv:2510.01060},
        pages = {arXiv:2510.01060},
          doi = {10.48550/arXiv.2510.01060},
archivePrefix = {arXiv},
       eprint = {2510.01060},
 primaryClass = {physics.plasm-ph},
       adsurl = {https://ui.adsabs.harvard.edu/abs/2025arXiv251001060V}
}

@ARTICLE{Werner2016ApJL,
       author = {{Werner}, G.~R. and {Uzdensky}, D.~A. and {Cerutti}, B. and {Nalewajko}, K. and {Begelman}, M.~C.},
        title = "{The Extent of Power-law Energy Spectra in Collisionless Relativistic Magnetic Reconnection in Pair Plasmas}",
      journal = {\apjl},
         year = 2016,
        month = jan,
       volume = {816},
       number = {1},
          eid = {L8},
        pages = {L8},
          doi = {10.3847/2041-8205/816/1/L8},
archivePrefix = {arXiv},
       eprint = {1409.8262},
 primaryClass = {astro-ph.HE},
       adsurl = {https://ui.adsabs.harvard.edu/abs/2016ApJ...816L...8W}
}

@ARTICLE{Werner2018MNRAS,
       author = {{Werner}, G.~R. and {Uzdensky}, D.~A. and {Begelman}, M.~C. and {Cerutti}, B. and {Nalewajko}, K.},
        title = "{Non-thermal particle acceleration in collisionless relativistic electron-proton reconnection}",
      journal = {\mnras},
         year = 2018,
        month = feb,
       volume = {473},
       number = {4},
        pages = {4840-4861},
          doi = {10.1093/mnras/stx2530},
archivePrefix = {arXiv},
       eprint = {1612.04493},
 primaryClass = {astro-ph.HE},
       adsurl = {https://ui.adsabs.harvard.edu/abs/2018MNRAS.473.4840W}
}

@ARTICLE{Xu2023ApJ,
       author = {{Xu}, Siyao and {Lazarian}, Alex},
        title = "{Turbulent Reconnection Acceleration}",
      journal = {\apj},
         year = 2023,
        month = jan,
       volume = {942},
       number = {1},
          eid = {21},
        pages = {21},
          doi = {10.3847/1538-4357/aca32c},
archivePrefix = {arXiv},
       eprint = {2211.08444},
 primaryClass = {physics.plasm-ph},
       adsurl = {https://ui.adsabs.harvard.edu/abs/2023ApJ...942...21X}
}

@article{Zhang2023,
title = {Particle acceleration in self-driven turbulent reconnection},
journal = {Journal of High Energy Astrophysics},
volume = {40},
pages = {1-10},
year = {2023},
issn = {2214-4048},
doi = {https://doi.org/10.1016/j.jheap.2023.08.001},
url = {https://www.sciencedirect.com/science/article/pii/S221440482300037X},
author = {Jian-Fu Zhang and Siyao Xu and Alex Lazarian and Grzegorz Kowal}
}
\bibliographystyle{aasjournalv7}



\end{document}